\documentclass[fleqn,usenatbib]{mnras}

\usepackage[T1]{fontenc}

\DeclareRobustCommand{\VAN}[3]{#2}
\let\VANthebibliography\thebibliography
\def\thebibliography{\DeclareRobustCommand{\VAN}[3]{##3}\VANthebibliography}

\usepackage{graphicx}	
\usepackage{amsmath}	
\usepackage{amssymb}	

\usepackage{xspace}

\usepackage{newtxtext,newtxmath}

\newcommand{\fL}{$f_{\mathrm {L}}$\xspace}

\newcommand{\fLt}{${f_{\mathrm {L}}{\rm[true]}}$\xspace}
\newcommand{\fLo}{${f_{\mathrm {L}}{\rm [obs]}}$\xspace}

\newcommand{\Npop}{${N_{\mathrm {pop}} }$\xspace}
\newcommand{\Nsamp}{${N_{\mathrm {samp}} }$\xspace}
\newcommand{\Npool}{${N_{\mathrm {pool}} }$\xspace}
\newcommand{\Dtel}{${D_{\mathrm {tel}} }$\xspace}

\newcommand{\Nt}{${N_{\mathrm {t}} }$\xspace}
\newcommand{\Tobs}{${T_{\mathrm {obs}} }$\xspace}
\newcommand{\Tmiss}{${T_{\mathrm {miss}} }$\xspace}

\title[The DRAKE mission]{The DRAKE mission: finding the frequency of life in the Cosmos}

\author[S. Sarkar]{
Subhajit Sarkar,$^{1}$\thanks{E-mail: subhajit.sarkar@astro.cf.ac.uk (SS)}
\\
$^{1}$School of Physics and Astronomy, Cardiff University, Cardiff CF24 3AA UK\\
}

\date{Accepted XXX. Received YYY; in original form ZZZ}

\pubyear{2020}

\begin{document}
\label{firstpage}
\pagerange{\pageref{firstpage}--\pageref{lastpage}}\maketitle

\begin{abstract} 
In the search for life in the Universe, exoplanets represent numerous natural experiments in planet formation, evolution, and the emergence of life.  
This raises the fascinating prospect of evaluating cosmic life on a statistical basis.  One key statistic is the occurrence rate of life-bearing worlds, \fL, the `frequency of life' term in the famous Drake Equation.  Measuring  \fL would give profound insight into how common life is and may help to constrain origin-of-life theories. I propose \fL as the goal for the DRAKE mission (Dedicated Research for Advancing Knowledge of Exobiology): a transit spectroscopy survey of M-dwarf habitable zone terrestrial planets. I investigate how the uncertainty on the observed value of \fL scales with sample size.  I determine that sampling error dominates over observational error and that the uncertainty is a function of the observed \fL value.  I show that even small sample sizes can provide significant constraints on \fL, boding well for the transit spectroscopy approach.  I perform a feasibility study of the DRAKE mission using a nominal instrument design and mission plan.  Due to low observing efficiencies, DRAKE may need to be incorporated into a wider-ranging deep-space or lunar observatory.  A 50-planet survey could constrain \fL  to $\leq$ 0.06 (at 95\% confidence) if the sample \fL = 0, or 0.03-0.2 if the sample \fL = 0.1.  This can be achieved (on average) in 10 years using a 17-m telescope with an unrestricted field-of-regard.  DRAKE is a viable approach to attempting the first experimental measurement of \fL.

\end{abstract}




\section{Introduction}

The search for life is one of the major goals of the field of exoplanet science. The past decade has seen great progress made in establishing the demographics of the exoplanet population with results that are encouraging for this search.  What has emerged is that small rocky planets greatly outnumber gas giants and that such planets appear to be common in the circumstellar habitable zones of a range of stellar types \citep{Petigura2013}.  Planets of this type are the focus in the search for water-based life akin to that of the Earth.  The frequency of such planets per star, $\eta_\oplus$, has been estimated for Sun-like and M-dwarf stars based on Kepler statistics. \cite{Petigura2013} considered Earth-sized planets, defined as between 1-2 $R_\oplus$, finding $\eta_\oplus$ of 22\% assuming that the habitable zone (HZ) is defined as between 0.25-4 $F_\oplus$ (where $F_\oplus$ is Earth-equivalent stellar irradiance).  Using the more conservative definition of the HZ by \cite{Kopparapu2013}, based on the moist greenhouse inner limit and
maximum greenhouse outer limit, they found  $\eta_\oplus$ falls to 8.6\%.  By comparison, for M-dwarf stars, using this same conservative HZ definition, \cite{Dressing2015} (see their table 8) found an $\eta_{\oplus}$ of 
27.36\% for planets between 1-2 $R_\oplus$ (15.85\% for 
1-1.5 $R_\oplus$ + 11.54\% for 
1.5-2 $R_\oplus$), indicating a higher HZ occurrence rate for terrestrial planets around M-dwarfs than around Sun-like stars.  

These planets represent innumerable natural experiments in the formation of Earth-like worlds and possible origins of life. As such, astrobiological questions of a statistical nature can start to be addressed.  A fundamental question is how common is life in the Universe? One paramaterisation of this is represented in the famous Drake Equation \citep{drake_2015} by the term \fL, the occurrence rate of life-bearing worlds. In this paper, I discuss \fL as a potential observable for future spectroscopic surveys of habitable zone exoplanets.  I then present results from a bootstrap Monte Carlo simulation elucidating the relationship between the sample size and experimental uncertainty on observed \fL values.  I then discuss a dedicated transit spectroscopy space mission called DRAKE (Dedicated Research for Advancing Knowledge of Exobiology) and present results of a feasibility study. The DRAKE mission is designed to perform a survey of M-dwarf habitable zone terrestrial planets with the primary goal of elucidating \fL for this subset.  
 
\section{The frequency of Life}
\label{sec: freq life}

The Drake Equation gives a sequence of physical factors that in combination give an estimate for the number  of communicative civilisations, $N$, in the Galaxy:  
\begin{equation}
N =  R_{*}\cdot f_{\mathrm {p} }\cdot \eta_{\oplus }\cdot f_{\mathrm {L} }\cdot f_{\mathrm {i} }\cdot f_{\mathrm {c} }\cdot L . 
\end{equation}
The latter terms: $f_{\mathrm {i} }$ (the fraction of life-bearing planets developing intelligence), $f_{\mathrm {c} }$ (the fraction of these developing communication) and $L$ (lifetime of such civilisations) relate to development of intelligence and technology once life has emerged and are unlikely to have experimental verification in this era. However, the first four terms each represent a potentially measurable observable. The first three of these terms already have experimental measurements: $R_{*}$ (the rate of star formation) [e.g. 1.65 $M_{\odot}$/yr \citep{Licquia2015}], $f_{\mathrm {p}}$ (the fraction of stars with planets) [$\sim$ 1 \citep{Cassan2012, Batalha2014}], and $ \eta_{\oplus }$ (the number of planets per stellar system with an environment suitable for life, interpreted here as the occurrence rate of terrestrial planets in the habitable zone).

A survey of atmospheric spectra of terrestrial planets in the habitable zones of stars could potentially, for the first time, provide experimental measurements for the fourth term, \fL, the fraction of planets where life emerges.  I term this the `frequency of life' and define it here as the ratio of the number of terrestrial planets in the habitable zones of stars where life has emerged to the total number of such planets. As such \fL depends on the definition of the HZ boundaries as wider boundaries will fold in more `candidate' planets, diluting the \fL value. In addition, such a survey will only be sensitive to life `as we know it', which amounts to carbon-based biochemistries producing identifiable biosignature gases.  Life which might arise outside the habitable-zone would not be detected in such a survey.

The inner and outer boundaries of the HZ define a region within which liquid water could exist on the surface of a planet.  Frequently used definitions are those based on
1-D cloud-free planet models by \cite{Kasting1993} and later modified by \cite{Kopparapu2013}.  These studies define the inner `moist greenhouse' and outer `maximum greenhouse' limits, referred to as the  `conservative' HZ,  while inner `recent Venus' and outer `late Mars' limits define the `optimistic' HZ.  However, there exists much debate on the extent of the habitable zone.  The effects of clouds, reduced planetary rotation, or desert-like worlds can bring the inner limit closer to the star \citep[e.g.,][]{Kitzmann2010,Zsom2013, Yang2014}.  The effects of a H/He dominated atmosphere could extend the outer limit \citep{Pierrehumbert2011}, potentially into interstellar space when combined with internal heat sources \citep{Stevenson1999}.  
A future survey of potential habitable zone planets could itself provide experimental constraints on the habitable zone boundaries that would provide a firmer determination of $ \eta_{\oplus }$ and thus 
\fL.  For the purposes of this paper,  we assume the optimistic HZ boundaries of \cite{Kopparapu2013} as the basis for selecting candidate planets from which to measure \fL.

Measuring \fL will give insight into how common life is and how life correlates with different planetary conditions. The latter could potentially help to constrain origin-of-life theories.
The \fL factor also features in the `Biosignature Drake Equation' or Seager Equation \citep{Seager2018}.  The Seager Equation calculates the number of planets, $N$,  with detectable signs of life based on biosignature gases:
\begin{equation}
N =  N_{*}\cdot f_{\mathrm {Q} } \cdot \eta_{\oplus }  \cdot f_{\mathrm {O}  } \cdot f_{\mathrm {L}} \cdot f_{\mathrm {S} },
\end{equation}
where $N_{*}$ is the number of stars considered, $f_{\mathrm {Q} }$ is the fraction of these stars that are suitable for planet-finding (quiescent, non-variable or non-binary),  $\eta_{\oplus}$ is the occurrence rate of HZ terrestrial planets, $f_{\mathrm {O}} $ is the fraction of such planets that are observable (by way of planet orbital geometry),
$f_{\mathrm {L}}$ is the fraction of planets that have life, and $f_{\mathrm {S}}$ is the fraction for which detectable biosignature gases are produced.
 We can rearrange this equation to solve for $f_{\mathrm {L}}$:
\begin{equation}
f_{\mathrm {L}} = \left( \frac{N} { N_{*}\cdot \eta_{\oplus }  \cdot f_{\mathrm {O} } } \right) 
\left(  \frac{1}{f_{\mathrm {Q}} \cdot f_{\mathrm {S} }} \right)   
\end{equation} 
Here the first bracketed term represents the ratio of the number of planets in a sample positive for biosignatures in the atmospheric spectrum to the total sample of planets observed.  Assuming this sample is representative of the larger population, this ratio will give an observed sample value for $f_{\mathrm {L}}$.  This value does not account for possible false negatives captured in the second bracketed term, e.g. planets orbiting stars that impact the detection of biosignatures or planets where life exists but detectable biosignature gases are not produced.  Constraining $f_{\mathrm {S}}$ is highly challenging in this era as we would need to detect life without the presence of biosignature gases.  Theoretical work has been done looking at the impact of UV radiation on M-dwarf Earth-like planet spectra \citep{Rugheimer2015} which may go some way to constraining $f_{\mathrm {Q} }$, however a definitive value for this term is not yet available.  As such, in this paper I consider $f_{\mathrm {L}}$ only for planets where life is detectable through biosignature gases and without accounting for the potential impact of the host star on the biosignature detectability.

The false negative rate might also include life that arises from alternate non-carbon-based biochemistries (e.g. silicon-based)
\citep{Petkowski2021, Bains2004} that might generate unrecognisable biosignatures.
The habitable zone concept is based around liquid water as the medium for biochemical reactions and carbon-based organic chemistry works well in water.  This together with favorable  physico-chemical properties of water \citep{Pohorille2012} for mediating macromolecular interactions and the high cosmic abundance of carbon together with its ability to easily form complex molecules \citep{Schwieterman2017} may strongly favour the paradigm of water- and carbon-based life in the habitable zone planets we are surveying.  It may be the case that such alternate biochemistries become more important under different physico-chemical conditions with non-water-based solvents \citep{Bains2004} as might be present in planets or moons outside the habitable zone. If so, then the false negative rate from this effect for planets within the habitable zone might be small, however this and the other 
 false negative considerations might lead to an unquantified bias in estimating the true $f_{\mathrm {L}}$ which is an unavoidable limitation of this study.
 
\section{The DRAKE mission}
\label{Drake}

A survey of terrestrial planets in the habitable zones of stars may not only provide the first constraints on \fL but could address other fundamental questions.   What is the range of planetary and atmospheric compositions in the HZ?  What initial and boundary conditions determine if a planet evolves onto an `Earth-like' track as opposed to `Venus-like' or `Mars-like' tracks? Are these clearly distinct evolutionary paths or more of a continuum of outcomes?  Can the presence of surface liquid water be detected and thus experimental constraints to the habitable zone determined? 
 
While the Hubble WFC3 has detected water vapour around super-Earth and sub-Neptune planets, its wavelength coverage is narrow (e.g. the G141 grism extends from 1.1-1.7 \textmu m), which limits its capability to detect biosignatures.   The first realistic prospects for biosignature detection will come through transit spectroscopy with JWST. However, JWST will still require many tens to hundreds of stacked transit observations to characterise a small number of the highest signal-to-noise habitable zone rocky planets \citep{Lustig-Yaeger2019, Barstow2015, Barstow2016, Greene2016, Seager2009}.  Ground-based high-dispersion spectroscopy (HDS) is another technique proposed to detect molecular oxygen biosignatures in terrestrial planets.  Simulations of HDS on the E-ELT have indicated the oxygen 0.76 \textmu m `A-band' could be detectable on Earth-like planets around nearby M-dwarf stars \citep{Snellen2013, Rodler}. However, the number of transit observations required would take many years to complete in each case and would be sensitive to only close-by and later M-dwarf hosts.  Combining HDS with high contrast imaging (HCI) has also been proposed to further improve detectability of both Earth-like planets and  of oxygen in their atmospheres. \cite{Snellen2015} simulated HDS/HCI with a hypothetical optical integral field spectrograph for the ELT METIS instrument finding that an Earth-sized planet in the habitable zone of Proxima Centauri could be detectable in one night of observing.  \cite{Lovis2017} simulated HDS/HCI coupling the SPHERE high-contrast imager and the ESPRESSO spectrograph on the VLT, finding that oxygen could be detected at 3.6$\sigma$ in 60 nights of observation spread over 3 years. Thus detecting biosignatures with HDS, with or without HCI, remains challenging, requiring many years of observing time and may be limited to nearby M-dwarfs and thus a small overall sample. Current capabilities to to perform a large survey of Earth-like planet spectra are therefore limited.  For such a survey, we must ideally develop a new facility dedicated to achieving that specific goal.

Previous mission concepts to obtain large numbers of Earth-like planet spectra have focused on direct-imaging  techniques with space-based formation-flying nulling interferometers such as Darwin \citep{Cockell2009}, Terrestrial Planet Finder \citep{Beichman1999} and LIFE \citep{Quanz2018}. 
While this approach has many advantages (e.g. a wide range of possible stellar host stars, and large potential sample sizes), significant technological development is still needed.
 
Instead, I propose the DRAKE mission concept based on the well-established technique of transit spectroscopy \citep{Seager2000, Charbonneau2002}. The purpose of this mission is to survey a sample of habitable zone terrestrial planets, obtaining atmospheric spectra through transmission spectroscopy at sufficient signal-to-noise ratio (SNR) and wide enough wavelength coverage to detect and characterise biosignatures, with the prime goal of obtaining the first constraints on \fL.  The technological development required would be significantly less than for direct imaging methods, and it would benefit from on-going community efforts to optimise the data from existing transit spectroscopy campaigns. 

There are several reasons why transit spectroscopic approaches might not have been previously favoured.  
Transit spectroscopy of terrestrial HZ planet atmospheres is essentially limited to  M-dwarf stars.  Due to their high molecular weight atmospheres, low temperatures and small radii, the spectral amplitudes observed in transmission spectra from such planets are much smaller than for hot Jupiters. These amplitudes are enhanced as the stellar radius falls, making their detection over noise more feasible around M-dwarf stars than Sun-like stars. This is also true in emission, where the planet-star flux ratio will be greater around M-dwarfs than Sun-like stars. Even so, co-addition of several transit or eclipse observations is required to reduce the noise, and this becomes impractical except around M-dwarfs where the periods in the HZ are in the order of days not years.  
This well-known `M-dwarf advantage' essentially rules out study of Earth-like planets around Sun-like stars using transit spectroscopy.  

The habitability or biosignature detectability of M-dwarf planets may be affected by stellar flares as well as X-ray and UV radiation from such stars \citep[e.g.][]{Wheatley2016, Rugheimer2015}. 
In additional star spots can impact the   accurate interpretation of a transmission spectrum.  Both occulted and unocculted star spots can bias transit depth measurements if not adequately corrected for \citep{Pont2013}. This may require a correction at the level of the light curve for occulted spots or a wavelength-dependent correction to the transmission spectrum for unocculted spots \citep[e.g.][]{Sing2011}.  The spot filling factor is needed for the latter correction and can be estimated through monitoring of the stellar variability.  In M-dwarfs, the contamination signal from unocculted spots may be significant and corrections may be complicated by the presence of faculae and potential uncertainties in determining the filling factor from variability measurements \citep{Rackham2018}.  Spot correction efforts remain an area of active research and could potentially utilise additional techniques such as doppler imaging to constrain the filling factor.
Stellar pulsations and granulation produce a correlated noise component on the light curve.  This may require decorrelation if the photon noise levels are very low especially at shorter wavelengths where it is more prominent, although the absolute level of this noise may be lower in M-dwarfs than Sun-like stars \citep{Sarkar2018}. 

The evolutionary development of planets and their atmospheres around M-dwarfs may be different than around Sun-like stars, both due to the longer lifetimes of the stars and the different spectral energy distribution, which could result in more abiotic oxygen false positives  \citep[e.g.][]{Luger2015,Domagal-Goldman2014}.  There are several other considerations (e.g.  IR photosynthesis, stellar variability, tidal locking) \citep{Heath1999, Tarter2007} which make it clear that life around an M-dwarf must be considered a particular subset of all possible life, and so extrapolation of results to the wider galactic basis must be carefully evaluated.  However, an alternate argument is that M-dwarf HZs are actually the galactic norm, with M-dwarfs hosting the majority of planetary systems, and having a higher occurrence of rocky planets in their HZs than Sun-like stars.  As such, they could be considered a priority for such studies.

Another limitation is the geometric transit probability of habitable zone planets which  greatly reduces the available sample size.  This is increased for M-dwarfs compared to Sun-like stars but remains of the order of about 1\%. In addition such M-dwarf systems would need to be bright (thus close), to minimise the fractional photon noise.

Until recently, there was a paucity of known terrestrial planets ($\leq$ 2 $R_\oplus$) in the habitable zones of M-dwarfs, and even now the numbers of such planets around close M-dwarfs, amenable to spectroscopic follow-up, remain low.  At this time 59 planets of radius $\leq$ 2.5 $R_\oplus$ are known within the optimistic habitable zones boundaries of \cite{Kopparapu2013} \footnote{http://phl.upr.edu/projects/habitable-exoplanets-catalog}.  Of these, 14 are < 2 $R_\oplus$ in size and transiting M-dwarfs.  Only 9 of these 14 are within
  100 pc. The five closest of these, TRAPPIST-1 planets d,e,f and g, and LHS 1140 b, have had transmission spectra obtained using the Hubble Wide-Field Camera 3 \citep{dewit2018, Edwards2021}, with a possible detection of water vapour on LHS 1140 b.  The number of known HZ planets orbiting bright M-dwarfs will continue to increase in the coming years as a result of the Transiting Exoplanet Survey Satellite (TESS) mission, together with the CHEOPS satellite, the future PLATO mission and ground-based surveys. 
In terms of the TESS yield for planets of < 2 $R_\oplus$ in the optimistic HZ of M-dwarf stars,
simulations by \cite{Barclay2018} predicted nine such planets, while \cite{Sullivan2015} predicted 14$\pm$4.  The extended TESS mission is likely to increase these predicted yields \citep{Bouma2017}. However, these projections indicate that the total number of known transiting Earth-sized planets within the optimistic HZ boundaries of M-dwarfs may remain in the order of tens rather than hundreds for the foreseeable future.  
A more liberal definition of the habitable zone brings a larger yield.  If the habitable zone is defined as 0.2-2 $F_\oplus$, \cite{Sullivan2015} predict that 48$\pm$7 planets of < 2 $R_\oplus$ will be discovered.  If it can be shown that small sample sizes of the order of a few tens can provide significant constraints on \fL, then the DRAKE mission's goal of 
finding \fL  becomes a viable prospect.

\section{Sample size}
\label{section: Effect}

It is thus important to establish the relationship between sample size and experimental uncertainty on any observed value for $f_{\mathrm {L}}$.   In this section, I attempt to establish this relationship using a bootstrap Monte Carlo simulation. This is also used to predict the minimum sample size for a single detection for any given population \fL.

The overall uncertainty in the observed value of \fL will result from the uncertainty in positive biosignature identification, observational error (due to noise on the detection), and sampling error due to the sample size.
The uncertainty in identification is a complex issue that requires delineating all possible gaseous biosignatures, the subject of much current research and debate. Earth-like biosignatures include the products of metabolism such as oxygen (O$_2$) and its by-product ozone (O$_3$), methane (CH$_4$) and nitrous oxide (N$_2$O) (discussed further in Section \ref{sec: Spectrometer}) \citep[e.g.][]{DeMarais2002, Segura2003, Kalt2009, Seager2014, Schwieterman2017}.
However, biosignatures may present differently at earlier times in a planet's evolution \citep[e.g.][]{Kalt2007} or if non-Earth-like biochemistries have evolved.  False positives for oxygen in particular must be accounted for, as discussed later.  

For this study, I have assumed that, in the absence of any noise on the spectrum,  positive or negative biosignature identification can be made with certainty on each spectrum.  While this simplification might not be justified with our current knowledge, the database of spectra collected on any survey could be re-analysed repeatedly into the future  and, as our understanding of biosignatures improves, such identification can be made with growing certitude.  I therefore examine only the effects of observational error and sampling error on the overall observed \fL uncertainty in relation to sample size.

In the following discussion I term the observed sample value of \fL as \fLo, i.e. the measured experimental value of \fL found from a survey of habitable zone terrestrial planets of sample size \Nsamp.  The true population value of \fL is termed \fLt, i.e. the actual value of \fL in the population of such planets.  The primary goal of the simulation is to find how the uncertainty or experimental error on \fLo scales with sample size, \Nsamp. This is parameterised as the 95\% confidence interval (CI), which is defined here as the range of most probable  values of  \fLt that could give that value of \fLo 95\% of the time: a measure of the likely range of the true, unknown parameter.

\subsection{Monte Carlo simulation}

The simulation is run under two conditions. 
In the first condition, we ignore any observational error in assigning biosignature positivity due to noise on the final spectrum, so that the uncertainty on \fL is purely due to sampling error. In the second, we incorporate observational error as well as sampling error.
As mentioned above, I assume that there is no uncertainty in identifying a given pattern of gaseous signatures in a noiseless spectrum as either positive or negative for life.  

The simulation is initiated by setting up a total population of transiting habitable zone terrestrial planets of size \Npop.  If we consider M-dwarfs only, these make up $\sim$70\% of the $\sim$10$^9$ stars in the Galaxy. If we assume about half of these have a planet of 2 $R_\oplus$ or smaller in the habitable zone, then assuming a transit probability of about 1\%, this gives an order of magnitude for \Npop of $\sim 10^6$.  I thus adopt \Npop = $10^6$ in these  simulations.

A given sample size, \Nsamp, is chosen, where 
\Nsamp = 10, 20, 30, 40, 50, 100, 200, 500 or 1000 planets.   A value for \fLt is chosen, ranging from 0 to 1 in steps of 0.001. For each \fLt the following algorithm is implemented.
 Each planet in the population is randomly assigned to be either positive or negative for spectral biosignatures, while ensuring that the the total number of positive planets is exactly \Npop $\times$ \fLt. 
A sample of planets of size \Nsamp is randomly selected (without replacement) from this population repeatedly 1000 times.  In each sample, the 
 \fLo is recorded by taking the ratio of positive planets in the sample to the total sample size.
This way, 1000 measurements of \fLo are taken for each \fLt.  When accounting for the observational error, I assume that the spectrum from each planet
is identified as positive or negative at 3$\sigma$ significance. As such, I assume that there is a 99.73\% chance that the planet is correctly identified and a 0.27\% chance it is incorrectly identified. Based on these probabilities the assignment of each planet to positive or negative for life is either retained (correctly identified as a true positive or true negative) or flipped (to give a false positive or false negative due to noise on the observation).  
The simulation is repeated 10 times and an average set of results used for the subsequent analysis.

I use the occurrence rates of \fLo for each \fLt (averaged over the 10 simulations)  to estimate the minimum sample size required to obtain at least one positive planet at 95\% confidence.  This is a function of \fLt.  At each \fLt, the fraction of outcomes where \fLo is non-zero to the total number of outcomes, gives the `positive fraction', the probability that at least one planet in the sample gives a positive result.  The positive fraction is found at each \Nsamp and then a line fit is used to determine the \Nsamp value that corresponds to a positive fraction of 0.95, i.e. the sample size where there is a 95\% chance of at least one positive result.  The line fit is achieved using a polynomial in log-log space\footnote{I find that the following degrees of polynomial, $r$, give good fits to the region around the 0.95 positive fraction in the following \fLt ranges: $r=5$ if \fLt< 0.01, $r$ = 6 for 0.01 $\leq$ \fLt < $0.1$, and $r$ = 7 for \fLt $\geq$ 0.1.}.
In some cases, a line fit is not needed, e.g. if the positive fraction is already >0.95 with \Nsamp = 10 or if the positive fraction does not reach 0.95 even at \Nsamp = 1000 (in which case no result is recorded). The latter occurs when \fLt reaches very low values.

 \begin{figure}
 \begin{center}
 	\includegraphics[trim={0.1cm 0.1cm 0.1cm 0.1cm}, clip, width=1.0\columnwidth]{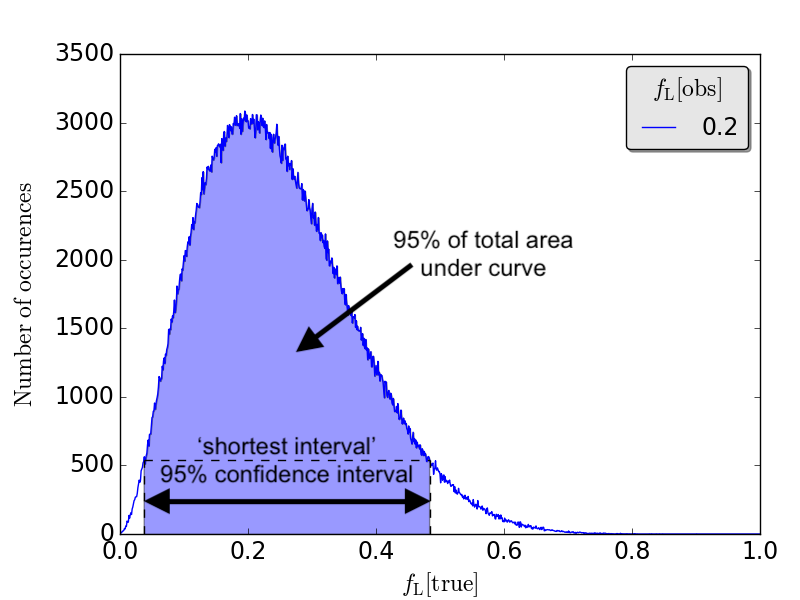}
 \end{center}
    \caption{  Shortest 95\% confidence interval.  This is the shortest distance parallel to the x-axis that contains 95\% of the area under the curve, with a combined tail probability of 5\% asymmetrically distributed between the two tails. This example shows the 95\% CI for an observed value \fLo = 0.2 when \Nsamp = 10 (based on simulations that included sampling and observational error and
    averaged over 10 simulations).
    The curve plots the number of occurrences of a given \fLt that resulted in an observed \fL of 0.2.  
    This distribution is proportional to the probability distribution and hence used to find the confidence interval as shown. }
\label{fig:shortest_interval}
\end{figure}

Next, we invert the occurrence rates to find the number of occurrences of \fLt for each \fLo.
The relative occurrence rates reflect the probability distribution of \fLt for a given \fLo and \Nsamp.  I use these distributions to find 95\% confidence intervals for each \fLo at each \Nsamp.  The range of possible \fLo values is a function of \Nsamp, varying from 0 to 1 in steps of  1/\Nsamp.  
 Since the distributions are in most cases skewed,  the `shortest interval' definition of the confidence interval \citep{Hall1988} is used, where there is a combined tail-end probability of $\alpha$, where, in this case, $\alpha$ = 0.05. This means that if \fLo is 0 or 1, the confidence interval (CI) will be entirely one-sided, which is appropriate since \fLt cannot be $<0$ or $>1$. 
 In each case, the shortest confidence interval is found by determining the shortest distance parallel to the x-axis between two points on the curve (of number of occurrences vs \fLt) that encloses 95\% of the total area under the curve\footnote{This is achieved using a custom code utilising the scipy functions: `interpolate' and `integrate'.} as shown in Figure \ref{fig:shortest_interval}.

\subsection{Results} 
\label{results 4.2}

\begin{figure*}
 \begin{center}
 	\includegraphics[trim={0cm 0cm 0cm 0cm}, clip, width=1.0\columnwidth]{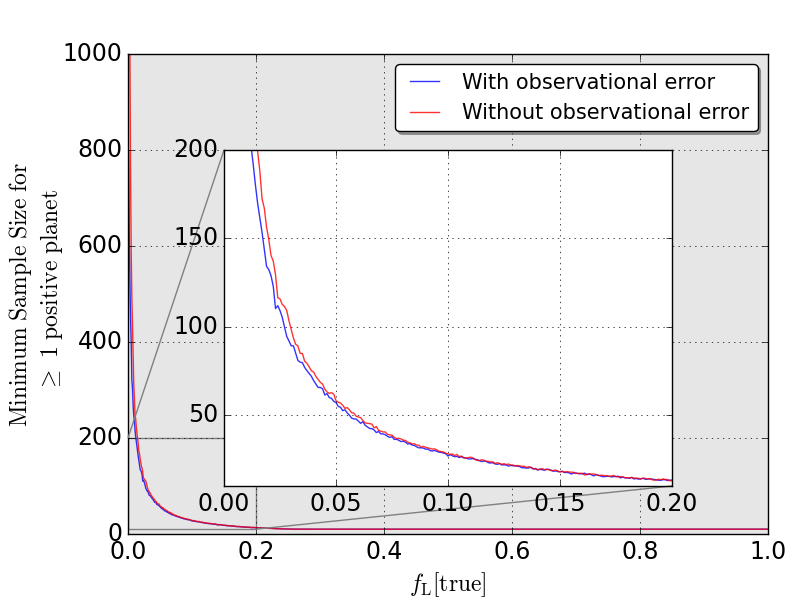}
 \end{center}
    \caption{Minimum sample size that returns at least one planet positive for biosignatures vs \fLt.  }
\label{fig:min_samp}
\end{figure*}
 
\begin{figure*}
 \begin{center}
 	\includegraphics[trim={0cm 0cm 0cm 0cm}, clip, width=1.0\columnwidth]{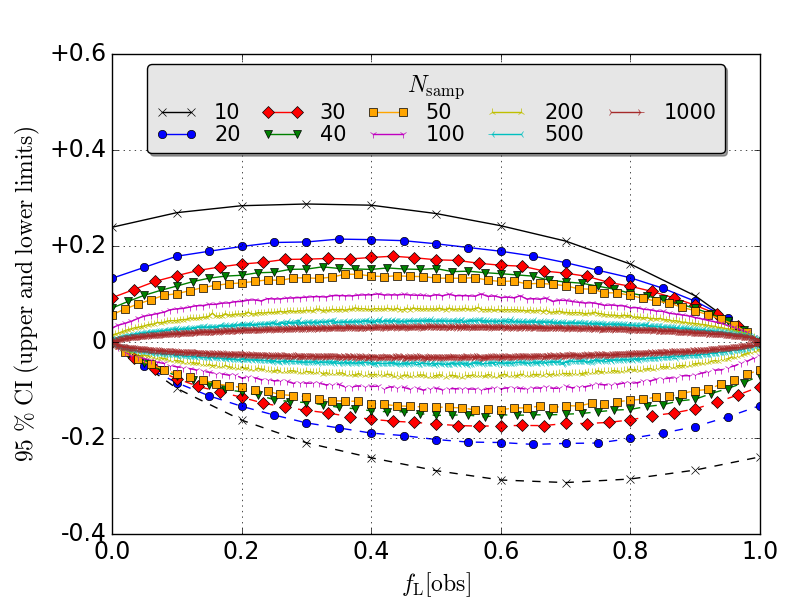}
 	\includegraphics[trim={0cm 0cm 0cm 0cm}, clip, width=1.0\columnwidth]{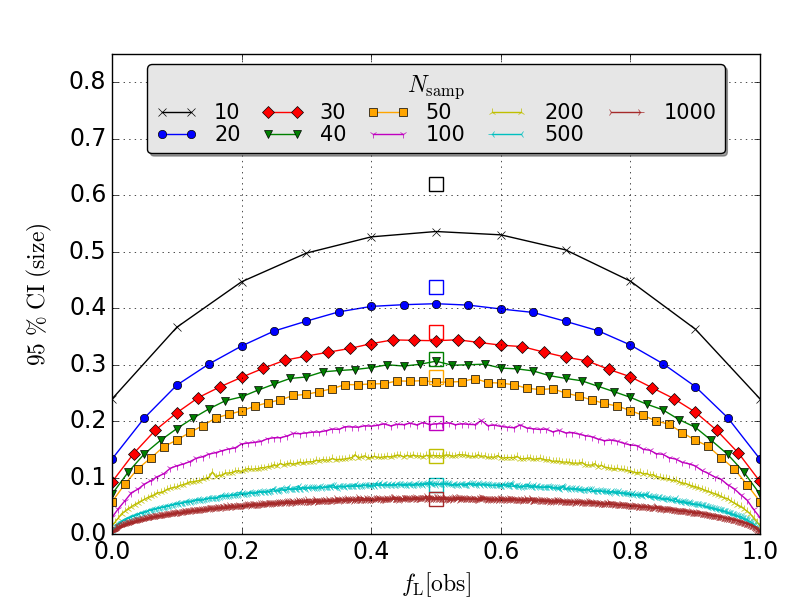}
 \end{center}
    \caption{95\% confidence intervals for \fLo for different sample sizes, \Nsamp. Left: CI upper (solid lines) and lower (dashed lines) limits. Right: CI absolute size. Also shown on the right plot as squares are the predicted maximum 95\% CI sizes using the Cochran formula (see text for details).}
\label{fig:CI1}
\end{figure*}

\begin{figure*}
 \begin{center}
 	\includegraphics[trim={0cm 0cm 0cm 0cm}, clip, width=1.0\columnwidth]{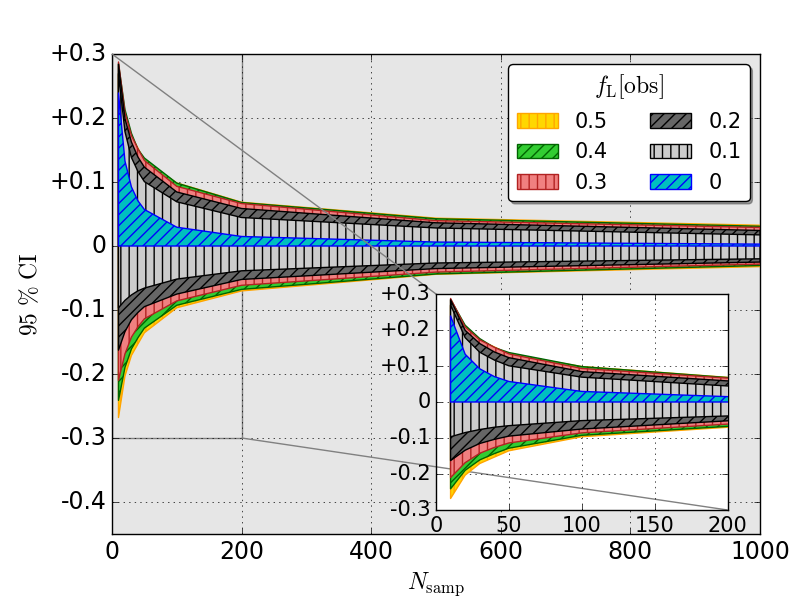}
 	\includegraphics[trim={0cm 0cm 0cm 0cm}, clip, width=1.0\columnwidth]{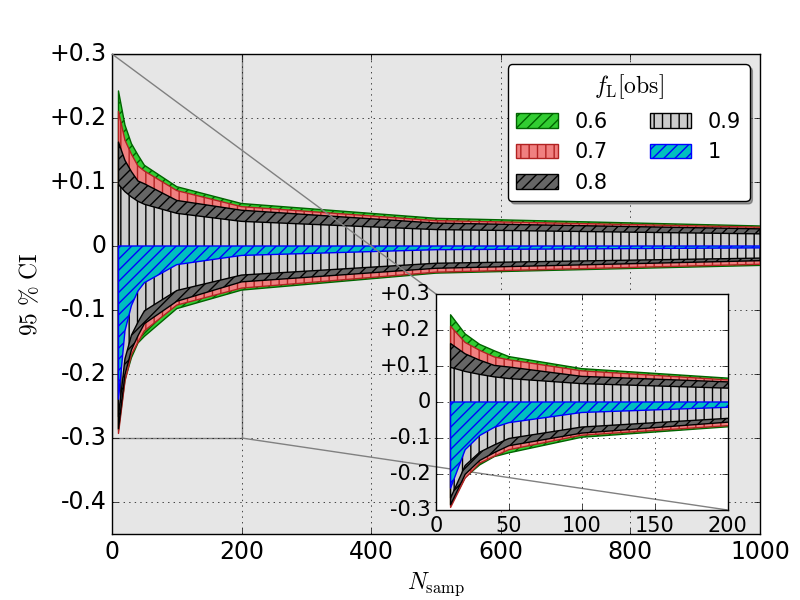}
 \end{center}
    \caption{95\% confidence interval vs \Nsamp for different \fLo values. Left: \fLo 0-0.5. Right: \fLo 0.6-1.}
\label{fig:CI2}
\end{figure*}

Figure \ref{fig:min_samp} shows the minimum sample size that will give at least one positive result vs \fLt.  Adding the observational error reduces the minimum sample size (the effect being greater at smaller \fLt) due to the inclusion of false positives.  As would be expected, larger \fLt corresponds to smaller minimum sample sizes. If \fLt were 5\% or more, then a sample size of 58 would give at least one positive result that was not a false positive due to observational error (Figure \ref{fig:min_samp}, inset).  This is similar to the prediction from \cite{LUVOIR2019}: 54 planets for one detection if the frequency of habitable planets is 5\%.
The problem with using this type of information is that we have no a priori information on \fLt and, as such, this kind of analysis is of limited use in designing a mission where the available number of planets might be small.  

The analysis of the uncertainty in \fLo might be more useful in this regard, as we make no assumptions on the underlying value of \fLt.  Due to the low probability assumed for false positive and false negative detections, which I have taken to constitute the observational error, the final results for \fLo uncertainty are dominated by sampling error, and as such no significant differences are found between the simulation that included observational error and the one that omits it.  Therefore I present and discuss here only the results that included the observational error.
Figure \ref{fig:CI1} (left) shows the upper and lower boundaries of the 95\% CI on \fLo vs \fLo for different sample sizes, \Nsamp.  These are asymmetric around the observed value except if \fLo = 0.5, and completely one-sided at \fLo of 0 or 1.  For example, if a survey had a sample of 20 planets, and the experimental result was an \fLo of 0.2, \fLt ranges around the observed value from -0.13 to +0.2, and the overall size of the CI is 0.33.  Figure \ref{fig:CI1} (right) shows the absolute size of the CI  vs \fLo for different \Nsamp values.  We see that the CI size is a function of \fLo and is maximal when \fLo = 0.5 and minimal at \fLo of 0 or 1.  Larger \Nsamp values result in smaller uncertainties, however, the incremental improvement in uncertainty is greatest at the smaller values of \Nsamp, so that pushing the sample size up has less and less impact on the final precision.  

To better visualise the relationship between \Nsamp and uncertainty, in Figure \ref{fig:CI2} the CI is plotted vs \Nsamp for different \fLo with the interval filled between the upper and lower limits.
We again see the value of increasing the sample size at small sample sizes and the reduced impact at large sample sizes.   
An interesting outcome of this study is the result for \fLo = 0, i.e. a completely negative sample.  Even in such a situation we can determine an upper limit to \fL at 95\% confidence.  The upper limit depends on the sample sized used, falling with \Nsamp.  If we consider a sample with just 10 planets returning an \fLo of 0, we can still constrain the upper limit of \fLt to 0.24 at 95\% confidence (Figure \ref{fig:CI2}, left, inset), i.e. there is a 5\% chance it would be higher than this.  While this is still a large uncertainty, it might still give a profound insight into the occurrence rate of life in the Universe.  If a sample size of 50 planets returned \fLo of 0, the upper limit to \fLt becomes 0.06, thus constraining the frequency of life to 6 percent or less at 95\% confidence (again not accounting for false negatives).  The same sample size would constrain \fLt to between 0.36-0.63 at 95\% confidence if an \fLo of 0.5 was measured.
 
Formulae such as Cochran's formula \citep{Cochran63} estimate the sample size, $n_0$, required to measure the fraction of a parameter, $p$, (e.g. \fL) in a large population to obtain a given margin of error, $e$, (due to sampling error) at a given confidence level (represented by the $Z$-score): $n_0  = Z^2p(1-p)/e^2$.
One reason it is not directly applicable here is that it returns the error on a measurement assuming knowledge of the underlying true value of $p$. For example, the Cochran formula would return an error of zero for $p$ = 0 or 1.  However, clearly the error on an observed value of \fL that is 0 or 1 will not be zero, so Cochran's formula cannot be directly applied in these cases, since we have no a priori knowledge of the underlying true population \fL.   If the true fraction of the parameter were 0.5, then the Cochran formula gives the maximum possible margin of error.  We find that for large values of \Nsamp, the maximum CI size on \fLo matches the prediction from the Cochran formula  well (Figure \ref{fig:CI1}, right). However, there is a deviation from the formula at low \Nsamp values where our method returns smaller CI sizes to the formula.  This may be related to progressive deviation of \fLt occurrence rates from the normal distribution as the sample size falls. 

We therefore find that even small sample sizes can return interesting results, with diminishing return on the improvement in uncertainty as the sample size is pushed up. A high-quality survey of about 50 planets could provide significant results for \fL.  It is plausible that in the next decade we may have such an available sample amenable to transit spectroscopy.


\section{Observatory design}
\label{sec: observ}

The DRAKE mission concept is centred around a transit spectroscopy instrument on a space-based visible-IR observatory that conducts a scheduled sequence of spectroscopy transit observations of the exoplanet sample.  Space is favoured over ground for transit spectroscopy: the noise is much lower due to the absence of atmospheric scintillation and turbulence, together with reduced observatory emission at longer wavelengths.  Furthermore, ground-based continuous wide-wavelength coverage is curtailed due to telluric absorption bands.  

I assume here an observatory design specially tailored for such a mission, however the goals of the DRAKE mission could equally be achieved as a specific component of a general purpose observatory project.  Possible locations include deep space, such as the L2 Lagrange point or a location on the Moon, such as the Shackleton crater at the lunar South Pole \citep{Schneider2021,Eads2021}.   Each location has advantages and disadvantages.  Low-Earth orbit, while allowing for the possibility of maintenance missions, would be subject to interruptions by diurnal cycles, and thus is not ideal for this mission where long out-of-transit baseline observations are needed to reduce the noise on the transit measurement and allow for better characterisation of host star activity and variability.

Designing a free-flying observatory at L2 would benefit from the heritage of previous observatories, such as Ariel and JWST, but would also have a limited life-span as consumables such as coolant and propellant run out.  I consider 10 years to be a likely upper limit for the mission lifetime without maintenance.
Maintenance with human missions would be unlikely at this location so that 
minimising mission duration is essential. For missions at L2, thermal constraints limit the allowable range of solar aspect angle, with an instantaneous field-of-regard (FOR) that is restricted and changing with time.  This means that different planets would be observable at different times as the observatory orbits the Sun.  Since the mission relies on conducting sequential observations on exoplanets that may be located anywhere in the sky, this will impact on the efficient scheduling of these observations, increasing mission duration.  Novel designs could be considered to increase the instantaneous FOR and thus reduce the mission duration. For example, if a free-floating Sun-shade is used, the FOR would be almost unrestricted except for the region occulted by the shield. An alternate power supply to solar-power would however be needed or a design with solar panels that could extend beyond the shaded zone.  

A polar lunar crater could provide a continuously-shaded thermally-stable location where solar power could be transferred from illuminated areas at a remote distance.  The FOR would be permanently restricted to one half of the sky so that roughly only half the possible number of candidate planets could be observed.  However the observable sample would be be always in the FOR so that scheduling efficiency would be close to that for an unrestricted FOR.  A possible second telescope at the other pole could simultaneously cover the planets in the other hemisphere, further increasing efficiency.  Lunar telescopes however would require huge investments in infrastructure, and while plans exist for humans to return to the Moon in the coming decade, it is unclear how this will develop with time.  A lunar telescope could however be maintained indefinitely, potentially allowing for a mission duration extending beyond 10 years.

The nominal mission assumes obtaining transmission spectra in primary transit, rather than emission spectra from secondary eclipse.  I assume that signals in transmission will generally be stronger than in emission. For example, examining the model transmission spectrum for an Earth-like planet orbiting the M3 star AD Leo (3390 K, 0.39 $R_{\odot}$) from \cite{Meadows2017} (their Fig. 2), which includes photochemistry effects, we find
that a typical spectral amplitude is about 4 ppm. For the same planet and star, a dayside emission spectrum (assuming the spectral amplitude $\approx F_{\rm p}(\lambda)/F_{\rm s}(\lambda)$, where $F_{\rm p}$ is the flux from the planet and $F_{\rm s}$ is the flux from the star arriving at the telescope, and modelling both planet and star as blackbodies) gives a maximum contrast ratio of 2.8 ppm  at 11 \textmu m (the longest wavelength proposed in the prototype design), and below 8.5 \textmu m, the signal is always below 1 ppm.
 
 I next describe the key features of a prototype design which forms the basis of the instrument model in the subsequent feasibility study.
  
\subsection{Telescope and common optics}
The telescope and common optics (TCO) describe all optical elements in the light path preceding the DRAKE spectrometer instrument.  The key elements are the primary mirror and the sequence of smaller mirrors in the light path which affect the overall transmission and will contribute to the optical emission background.  The primary mirror diameter, $D_{\rm tel}$, is the key design parameter which is investigated in the feasibility study.  $D_{\rm tel}$ is varied from 10 to 50 m.
The total transmission (TCO transmission combined with the spectrometer transmission), $\nu$, is assumed to be 0.6.


\subsection{Spectrometer}
\label{sec: Spectrometer}

The baseline concept is for a visible to infrared low resolution dispersive spectrometer providing wide wavelength coverage to maximise the identification of biosignatures in the pattern of spectral features from an exoplanet atmosphere. The wavelength range is divided into two channels.  This allows each channel to be optimised for its particular wavelength range, e.g. detector characteristics, spectral dispersion, spectral resolving power etc.

\subsubsection{Channel A (0.6-5 \textmu m)}
A spectrometer channel ranging from visible to near infrared (NIR) wavelengths (0.6-5 \textmu m) would permit detection of key biosignature molecules as well as the characterisation of clouds, hazes, and Rayleigh scattering, and monitoring for stellar activity.  \cite{Meadows2018} present a possible flow chart (their Fig. 11)  for spectroscopic identification of a photosynthetic biosphere or `Archaen Earth' exoplanet based on the presence or absence of water (H$_2$O), O$_2$, O$_4$,  CH$_4$, carbon dioxide (CO$_2$) and carbon monoxide (CO). This  aims mainly to interpret any detection of O$_2$ in the context of other gases that may or may not support its biological origin.  Abiotic origin of O$_2$ could result from photolysis of water, especially in ocean-loss scenarios, which might present with extremely high O$_2$ levels marked by the presence of O$_4$ \citep{Luger2015, Schwieterman2016}.  It could also arise from photolysis of CO$_2$,   correlating with large CO$_2$ abundances and the presence of CO as a photolytic by-product \citep{Hu2012, Domagal-Goldman2014}.

The wavelength coverage of Channel A includes features from all these gases. There is strong O$_2$ absorption at 0.76 \textmu m ( `A-band'), with weaker features at  0.69 \textmu m (`B-band') and 0.63 \textmu m (`$\gamma$ band'), and O$_4$ features at 1.06 and 1.27 \textmu m \citep{Meadows2018}.   Strong CH$_4$ features occur at 1.65, 2.4 and 3.3 \textmu m \citep{Schwieterman2017}, and large amounts of CH$_4$ could be supportive of biogenic rather than photolytic origin of O$_2$ as CH$_4$ is a sink for photochemically generated O$_2$ \citep{Meadows2018}.
Strong CO$_2$ absorption occurs at $\sim$ 1.65, 2 and 4.2 \textmu m and CO at 2.35 and 4.6 \textmu m \citep{Meadows2018}.   The presence of water vapour would be a necessary but not sufficient marker for life as we know it.  In this range there are water absorption bands at 0.65, 0.7, 0.73, 0.8, 0.95, 1.1, 1.4, and 1.8–2.0 \textmu m \citep{Meadows2018}.  Organic sulphide gases   are produced by bacteria and fresh water green and blue-green algae \citep{Rasmussen1974, Cooper1987, Pilcher2003}.  The sulphide gases methanethiol (CH$_3$SH), dimethyl sulphide (CH$_3$SCH$_3$) and dimethyl disulphide (CH$_3$SSCH$_3$), together with methyl chloride (CH$_3$Cl) (another potential biosignature associated with numerous sources such as algae, plants and fungi \citep{Khalil1999, Harper1985, Yokouchi2002}), all have an absorption band between 3-4 \textmu m  \citep[][Fig. 6]{Schwieterman2017}.  Nitrous oxide (N$_2$O) is a biosignature gas generated by nitrogen-fixing bacteria and algae \citep{Sagan1993}. In this channel, N$_2$O has significant features at 3.7 and 4.5 \textmu m \citep{Schwieterman2017}.
Trivalent phosphorus compounds have recently been proposed as possible biomarkers of anerobic organisms, in particular phosphine \citep{Sousa-Silva2020} for which there has been a claim of detection on Venus \citep{Greaves2020}. Phosphine has strong bands around 2.7-3.6 \textmu m and 4.0-4.8 \textmu m \citep{Sousa-Silva2020}.
 
\subsubsection{Channel B (5-11 \textmu m)}
Wavelength coverage extending into the mid-infrared (MIR) may be desirable to resolve degeneracies resulting from overlapping molecular features as well as between gas abundances and temperature structure (in eclipse) and gas abundances and cloud coverage (in transit) \citep{Barstow2015}, although eclipse observations are not included in the baseline design.  Channel B covers the wavelength range between 5-11 \textmu m.  This range includes a water band at 6.3 \textmu m and a very strong O$_3$ feature at 9.6 \textmu m. O$_3$ is a photochemical byproduct of O$_2$ \citep{Meadows2018}. The upper bound of the wavelength range is chosen to include the full O$_3$ feature but is not extended to longer wavelengths as the benefit of covering longer wavelengths is outweighed by the negative impact on SNR. N$_2$O absorbs at 7.8 and 8.6 \textmu m and CH$_4$ at $\sim$7-8 \textmu m \citep{Schwieterman2017}.  In addition, organic sulphide gases have several absorption features in this range,  e.g. dimethyl sulphide  at 6–7 and 10 \textmu m \citep{Schwieterman2017}. The relevance of organic sulphide gases as biosignatures is determined in relation to ethane (C$_2$H$_6$), which has an absorption feature in this channel at 6–7 \textmu m \citep{Schwieterman2017}.  CH$_3$Cl also has features in this range at 7 and 9.7 \textmu m \citep{Schwieterman2017}.  Phosphine has absorption at 7.8-11.5 \textmu m \citep{Sousa-Silva2020}.  

\subsubsection{Detectors}

Different optimal detector technologies may be needed for each channel, e.g. mercury-cadmium-telluride (MCT) for Channel A or arsenic-doped silicon impurity band conduction for Channel B.  At this stage I do not specify detailed characteristics of the detectors, such as dark current, read noise, pixel full-well capacity or flat-field variations, since the feasibility study will assume a photon-noise-limited instrument. For both channels the detector quantum efficiency, $QE$, is assumed to be 0.8\footnote{This is consistent with quantum efficiency measurements for HxRG detectors \citep[e.g.][]{Blank2011, Mosby2020} and for experimental LWIR MCT detectors \citep{Cabrera2019}.
}, and the quantum yield\footnote{The internal quantum yield is the number of electron-hole pairs produced per photon absorbed.} is assumed to be unity. The final photon-converting efficiency, $\kappa$ = $QE \times \nu$, is then 0.48 (simplified here to be non-wavelength dependent). 

\subsubsection{Spectral resolving power}

The spectral resolving power $R$ (where $R$ = $\lambda/\Delta \lambda$) 
 of the dispersive element sets a limit to the width of spectral features that can be detected, with narrower features needing higher $R$. So each channel must have a minimum $R$ power that permits the detection of key spectral features in its wavelength range.
 
The detectability of a given spectral feature may depend both on its SNR per spectral bin (which falls with $R$) and also on the number of points (spectral bins) sampled across the feature (which increases with $R$). The complex balance between these two effects will influence the optimal $R$ power but this will require more detailed studies with both more defined instrumental modelling and spectral retrieval studies.  While the relationship between SNR and $R$ is straightforward for a photon-noise-limited instrument, when instrumental noise sources such as read noise are included, it becomes more complex.  For example, changing the $R$ power affects the linear dispersion and the overall length of the spectral trace, which in turn affects the number of pixels per resolution element and hence the read noise and dark current noise. Other considerations include the point-spread-function (PSF), which will also affect the $R$ power. The PSF size should be such as to be Nyquist-sampled by the detector but it is also affected by wavefront-error aberration that is related to the surface quality of the primary mirror. The requirements on the latter may strongly impact feasibility, particularly through overall cost. PSF size
will also influence the size of the extraction aperture and hence the amount of instrumental and background noise folded into each spectral element. The linear dispersion and PSF shape will also impact the saturation time on the detector for a given pixel full-well capacity, limiting the maximum brightness of the target that can be observed.  Future trade-off studies will be needed to find the optimal instrumental design, which maximises the detectability of the final spectrum.  For now the  $R$ power per channel is based  on previous studies.

\cite{Robinson2016} modeled a coronographic instrument and found the oxygen A-band detectable at at Earth-like abundances with $R = 70$.  An analysis by \cite{Brandt2014} found that the optimal resolution for an A-band O$_2$ detection varied from $R \sim$ 70 to many hundreds, depending on the instrumental noise model adopted.  Thus an $R$ of 100 is adopted for Channel A and it is assumed that this is likely to be adequate for detecting the O$_2$ A-band and other features in Channel A.
 
 In the MIR, ro-vibrational spectral bands for biosignature gases tend to be broader than at shorter wavelengths \citep[][Fig. 6]{ Schwieterman2017} so that lower $R$ powers can be  considered.  \cite{Meadows2018} suggest an $R$ of 10 could be used to scan for broad features such as water or O$_3$. However, to to give some margin to detect narrower features at the shorter wavelengths, an $R$ of 30 is adopted for Channel B.  This is consistent with the minimum $R$ power of the Ariel long-wave channel (AIRS channel 1) operating between 3.9 and 7.8 \textmu m \citep{Tinetti2018}.
 
\subsection{Spacecraft} 
 Assuming a deep-space location, the TCO and spectrometer will interface with the spacecraft service module which controls vital functions such as power, telecommunication, data handling, thermal control and attitude control.  A lunar location would require a modification of this interface suitable for a ground-based observatory.




\section{Feasibility study}
 
In this section, I simulate the nominal DRAKE mission and assess its capability to fulfil the primary goal of obtaining an experimental value for \fL.  The mission consists of a sequence of transit spectroscopic observations of a sample of \Nsamp planets producing a transmission spectrum for each planet extending from 0.6 to 11 \textmu m.  

The feasiblity study has the following steps, which are described in more detail below. A population of Earth-sized planets around M-dwarf stars is simulated using a model planet population generator, filtered and then ordered by photometric detectability.  100 such randomised realisations are generated. In each instance, for each candidate planet, the SNR 
to detect a `typical' spectral feature is found at each wavelength in each channel for a single transit, $SNR_1(\lambda)$.  The minimum $SNR_1$ is then used to calculate the
the required number of  transit observations, $N_{\rm t}$.  This is done using an instrument model of the DRAKE observatory with a primary mirror size, \Dtel, of 30 m.  The SNR results (and thus \Nt)   are then scaled for different values of \Dtel ranging from 10-50 m.  Samples sizes, \Nsamp, are considered ranging from 10-100 planets
A scheduling algorithm is used to find the total mission time, \Tmiss, to complete observations for all \Nsamp planets. This is done by first selecting an initial pool of planets, \Npool, which is 1.2 $\times$ \Nsamp and chosen in order of photometric detectability.
From the 100 realisations for each case, average results are obtained. We thus obtain relationships between $T_{\rm {miss}}$, \Nsamp and \Dtel from which feasibility is assessed.

\subsection{Instrument model}

The instrument model is based on the design presented in Section \ref{sec: observ} and assumes a photon-noise-limited regime with no other influences on photometric stability. As such, we assume that other noise sources and systematics can be mitigated either through design or in data reduction to levels well below the photon noise.  These include read noise, dark current noise, pointing jitter, detector non-linearity, persistence and 1/f noise, as well as astrophysical factors including stellar activity \citep{Rackham2018} and stellar pulsation and granulation \citep{Sarkar2018}.  Background noise contributions from optical surface emissions and zodiacal light are assumed to be negligible. For the former, this should be a reasonable assumption with passive or active cooling of optical elements  to about 50-70 K. However, at the extreme long wavelength end of Channel B background contributions such as zodiacal light noise could become significant \citep{Sarkar2020}.

\subsection{Model planet population}

\begin{figure}
 \begin{center}
 	\includegraphics[trim={0cm 0cm 0.2cm 0cm}, clip, width=1.0\columnwidth]{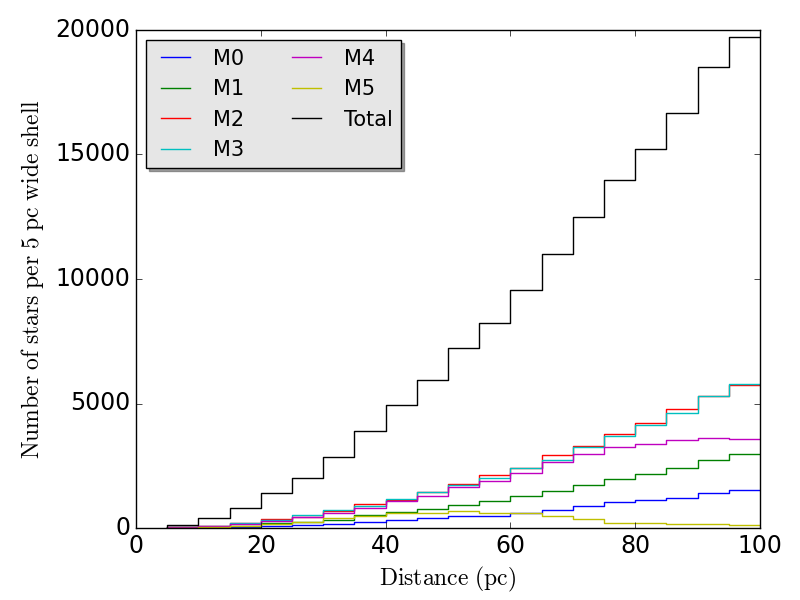}
 \end{center}
    \caption{M-dwarfs per subclass per 5 pc wide shell versus distance obtained from the
    TESS Input Catalogue \citep{Stassun2018} Candidate Target List 'cool dwarfs' special list that have TESS magnitude $\leq$ 16. }
\label{fig:TIC}
\end{figure}

We set up a model planet population for this study as follows.  The space around the Earth is divided into consecutive shells of width 5 pc out to a distance of 100 pc.  The TESS Input Catalogue Candidate Target List (CTL) \citep{Stassun2018} is mined to find the number of known M-dwarfs in each subclass within each shell\footnote{Data is obtained from https://filtergraph.com/tess\textunderscore ctl }. The search looks for stars included in the `cool dwarfs' special list with TESS magnitude of 16 or less.  Stellar subclass is assumed based on temperature, as categorised in \cite{Kalt2009}.  Using these criteria  no stars of M6 or later are found. Figure \ref{fig:TIC} shows the numbers of stars of each subclass in each shell. 

These distributions are used to set up a randomised model stellar population.  In each shell, stars of the same subclass are given the same radius ($R_{\rm s}$), mass ($M_{\rm {s}}$) and temperature ($T_{\rm {s}}$), based on \cite{Kalt2009}. However, in each realisation their distance, $d$, is randomised within the shell.  Each star is then randomly allocated a habitable zone planet or no planet.  If allocated a planet, it is specified as a `sub-Earth' ($0.8R_{\oplus}<R_{\rm p}<1R_{\oplus}$), an Earth-sized planet ($1R_{\oplus}<R_{\rm p}<1.5R_{\oplus}$), or a `super-Earth' ($1.5R_{\oplus}<R_{\rm p}<2R_{\oplus}$) based on the occurrence rates in \cite{Dressing2015} for optimistic (Early Mars - Recent Venus) HZ limits.  These are 13.09\% for sub-Earths, 24.28\% for Earth-sized planets, and 20.69\% for super-Earths.
Each planet is randomly assigned a semi-major axis, $a$, within the HZ boundaries for that star. The boundaries of the HZ are calculated for each star based on effective stellar flux values for Early Mars and Recent Venus boundaries adjusted according to Eq. 2 in \cite{Kopparapu2013}. A transit probability, $P_{\rm t}$, is assigned for each planet, where $P_{\rm t} = R_{\rm s}/a$. For each planet, if a random number between 0 and 1 falls below the planet's transit probability it is assumed to be transiting and is included for further consideration.  The best-fit spline model of planet size distribution from \cite{Fulton2017} (their Fig. 7) is used as a probability distribution for planet radius. For each transiting planet its radius, $R_{\rm p}$, is randomly assigned within the range of radii for its class (sub-Earth, Earth-sized or super-Earth), weighted by this probability distribution. 
The mass of each planet, $M_{\rm {p}}$, is assigned from
mass-radius relation of \cite{Otegi2020} where $M_{\rm {p}} = 0.9 R_{\rm p}^{3.45}$.
The equilibrium temperature for each planet, $T_{\rm {eq}}$, is calculated assuming it is not tidally-locked and has an albedo of 0.3.  

Each planet is assumed to have an atmosphere, with a mean molecular weight, $\mu$, which is assigned in two different ways.  In the first (type 1) $\mu$ is fixed to 29 amu\footnote{The approximate value for the Earth.} for all planets.  In the second (type 2) $\mu$ is randomly assigned between 15-45 amu\footnote{Effectively there are two sets of simulations performed: one for fixed $\mu$ and one for randomised $\mu$.}. The scale height of the atmosphere, $H$, is then given by $kT_{\rm {eq}}/(\mu g)$, where $g$ is the calculated surface gravity and $k$ is Boltzmann's constant.  The period, $P$, for each planet is calculated from Kepler's third law and the transit duration, $T_{14} = 2(R_{\rm p}+R_{\rm s})/v$, is calculated,  where $v$ is the orbital velocity.  Circular orbits are assumed, together with inclination angles of 90$^{\circ}$ in all cases. 

\subsection{Planet detectability}

Next, we must consider the fact that not all transiting planets that exist will be known. The list of known planets at the time of the mission will depend on their previous detection by transit photometric observing campaigns.  The following model is used to determine an order of relative detectability for each transiting planet in the model population.  

A hypothetical transit photometry survey is considered that scans the entire sky such that each star is observed continuously for the same total observing time, ${T_{\rm {obs}}}^{p}$ (where the superscript $p$ denotes photometer).  This is a simplification of how all-sky transit surveys actually perform where, in reality, different stars may be observed for different total observing times.
The SNR for detection, $SNR_{\rm {det}}$, is defined to be the ratio of the photometric fractional transit depth due to the planet, $\tau^p = (R_{\rm p}/R_{\rm s})^2$, to the noise on this transit depth, $\sigma_{\tau^p}$, and will differ for each planet. $\sigma_{\tau^p}$ will decrease with the square root of the number of transits observed (${N_{\rm t}}^p$) within the total observing time.  I assume  ${N_{\rm t}}^p = {T_{\rm {obs}}}^p/P$.  Each planetary transit observation is modeled as a `box-car'\footnote{Thus ingress and egress slopes are ignored, as is limb-darkening.} with an in-transit duration, $T_{\rm {14}}$, and an out-of-transit (OOT) duration, $P$-$T_{\rm {14}}$. A 100\% efficient duty cycle is assumed.  If the photometric observatory has a collecting area, $ {A_{\rm tel}}^p $, and photon conversion efficiency, $\kappa^p$, then assuming photon noise only (and $\tau^p << 1$), we can estimate $\sigma_{\tau^p}$ as:
\begin{equation}
\sigma_{\tau^p} \approx \left[{}T_{\rm obs}^p\kappa^p {A_{\rm {tel}}}^p\int^{\lambda_1}_{\lambda_0} F_{\rm tel}(\lambda)\frac{\lambda}{hc}d\lambda   \left(\frac{T_{14}[P-T_{14}]}{P^2} \right) \right]^{-1/2}
\label{Eq tau}
\end{equation}
where $h$ and $c$ are Planck's constant and the speed of light respectively, and $F_{\rm {tel}}(\lambda)$ is the stellar flux density received at the telescope, which is integrated over a wavelength passband ($\lambda_0-\lambda_1$).  
 $SNR_{\rm {det}}$ can then be estimated as the product of two factors: $\alpha$, a factor that is instrument- and observation-dependent, and $\beta$, a factor that depends on the planet and star parameters alone:
\begin{equation}
SNR_{\rm det} \approx \alpha \beta
\label{Eq det SNR}
\end{equation}
\begin{equation}
\alpha = \sqrt{\kappa^p  {A_{\rm tel}}^p  {T_{\rm obs}}^p }
\end{equation} 
\label{Eq alpha}
\begin{equation}
\beta= \left(\frac{R_{\rm p}}{R_{\rm s}} \right)^2  \left[\int^{\lambda_1}_{\lambda_0} F_{\rm tel}(\lambda)\frac{\lambda}{hc}d\lambda   \left(\frac{T_{14}[P-T_{14}]}{P^ 2} \right) \right]^{1/2}
\label{Eq beta}
\end{equation}
If we assume $\alpha$ is a constant for all planets for a given survey, then the relative detectability is given by $\beta$, which can be calculated for each planet.  I assume here that the wavelength passband ($\lambda_0-\lambda_1$) is between 0.5-1.0 \textmu m (since most photometric surveys use the optical range) and I approximate $F_{\rm {tel}}$ with a blackbody function so that $F_{\rm {tel}}(\lambda) = \pi B_{\lambda}(T_{\rm {s}})(R_{\rm s}/d)^2$.  
The relative photometric detectability is used to select planets for inclusion in the mission schedule in order of their detectability. For example, if 100 planets are to be included in the mission schedule, the 100 most detectable planets are chosen for inclusion.    
 
\subsection{Mission duration}

For $N_{\rm samp}$ = 10, 20, 30, 40, 50, 60, 70, 80, 90 and 100 planets and \Dtel = 10, 20, 30, 40 and 50 m,  $T_{\rm {miss}}$ is found using the following methodology.
 
\subsubsection{Atmospheric spectral feature detection}
For each planet we must first calculate the wavelength-dependent SNR for one transit, $SNR_1(\lambda)$, in each channel for detection of a `typical' atmospheric spectral feature targeted in each channel. 
The height and width of a typical spectral feature is estimated as follows.  

I utilise the publicly-available model transmission spectra produced by \cite{Wunderlich2019} for an Earth-analogue planet orbiting stars of different M-dwarf subclasses\footnote{http://cdsarc.u-strasbg.fr/viz-bin/qcat?J/A+A/624/A49}. These spectra include the impact of photochemistry and cover the full wavelength range of the DRAKE instrument.  They are also provided in units of effective height (in km) which allows conversion to dimensionless units of scale height.
These spectra were downloaded for the following M-dwarf subclasses: M1 (GJ 832), M2 (GJ 176, GJ 581,) M3 ( GJ 436, GJ 644, AD Leo, GJ 667C), M4  (GJ 876, GJ 1214) and M5 (Proxima Centauri). Each spectrum was resampled to standardised wavelength grid and divided into the Channel A and Channel B wavelength ranges.  In the Channel A range the following ro-vibrational spectral features were used to derive height and width: 0.9 \textmu m CH$_4$, 1.2 \textmu m CH$_4$/H$_2$O, 1.4 \textmu m CH$_4$/H$_2$O, 1.7 \textmu m CH$_4$, 2.3 \textmu m CH$_4$, 2.7 \textmu m CO$_2$/H$_2$O, 3.3 \textmu m CH$_4$, 4.3 \textmu m CO$_2$ and 4.7 \textmu m CO. In the Channel B range the following features were used: 5.9 \textmu m H$_2$O, 7.7 \textmu m CH$_4$ and 9.5 \textmu m O$_3$.  The central wavelengths given above are the weighted averages from the estimated extent of each feature.

The average peak effective height, $h_{\rm{peak}}$, for the features in each channel were determined for each spectrum per host star. These were further averaged for stars within each M-dwarf subclass, to give an average $h_{\rm{peak}}$ for each subclass (Table \ref{table: spec feat}).

Next, for each feature in each spectrum, the full-width half-maximum (FWHM) was estimated as follows.  In general, the boundary wavelengths of the feature at half-maximum ($\lambda_1$ and  $\lambda_2$) were found and this span taken to be the FWHM. In some cases (due to overlapping features) only one side of the spectral feature is able to yield a measurable boundary wavelength (e.g. $\lambda_1$ or $\lambda_2$ but not both).  In such cases, I take the difference between the above central wavelength and the measurable boundary wavelength and double it to get a FWHM.  For each channel, the average FWHM is obtained for each star, and then averaged again for each M-dwarf subclass.  Since the averaged FWHM estimates were similar across   the different subclasses, these were further averaged over all subclasses to give a single value for each channel: 0.3 \textmu m for Channel A and 1.2 \textmu m for Channel B\footnote{In a couple of cases (for GJ 644 and AD Leo) the FWHM of the 0.9 \textmu m CH$_4$ could not be estimated as neither side of the feature could yield a boundary wavelength due to the continuum and thus these did not contribute to the calculation of the average.}.

Next, I assume that the typical spectral feature is sampled at least twice over its FWHM to properly sample its shape, giving a maximum spectral bin size, $\Delta \lambda_A$, of 0.5 FWHM.  For each channel and M-dwarf subclass, the feature is modelled as a Gaussian function of height $h_{\rm{peak}}$ (Table \ref{table: spec feat}) and standard deviation = FWHM/2.355 (where the average FWHM estimates for Channel A and Channel B are used). Assuming the above spectral bin is centred over the peak, the average height within the  bin, $h_{\rm{bin}}$, is then calculated (and will be slightly lower than the peak value).  Since the planets in \cite{Wunderlich2019} are modelled as Earth analogues with  N$_2$-O$_2$-dominated atmospheres, I make the assumption that the scale heights are the same as for the Earth at 8.5 km.  Dividing $h_{\rm{bin}}$ in km by 8.5 km, we obtain $n_H$, the amplitude in units of the scale height (Table \ref{table: spec feat}). This method considers only a single spectral feature in isolation and does quantify the impact of adjacent spectral features, any gaseous continuum, clouds or haze.

\begin{table}
\begin{center}
\caption{Typical spectral feature estimates.}
\label{table: spec feat}
\begin{tabular}{ccccccc} 
\hline
\hline
\multicolumn{1}{c}{Subclass} &
\multicolumn{3}{c}{Channel A} &
\multicolumn{3}{c}{Channel B}
\\
\multicolumn{1}{c}{} &
\multicolumn{3}{c}{0.6-5 \textmu m} &
\multicolumn{3}{c}{5-11 \textmu m}
\\
\multicolumn{1}{c}{}  
&
\multicolumn{1}{c}{$h_{\rm peak}$}
&
\multicolumn{1}{c}{$h_{\rm bin}$}
&
\multicolumn{1}{c}{$n_H$} 
&
\multicolumn{1}{c}{$h_{\rm peak}$}
&
\multicolumn{1}{c}{$h_{\rm bin}$}
&
\multicolumn{1}{c}{$n_H$}
\\
\multicolumn{1}{c}{}  
&
\multicolumn{1}{c}{(km) }
&
\multicolumn{1}{c}{(km)}
&
\multicolumn{1}{c}{($H$)} 
&
\multicolumn{1}{c}{(km) }
&
\multicolumn{1}{c}{(km)}
&
\multicolumn{1}{c}{($H$)}
\\
\hline
\multicolumn{1}{c}{M1} &
\multicolumn{1}{c}{35} &
\multicolumn{1}{c}{33} &
\multicolumn{1}{c}{3.9} &
\multicolumn{1}{c}{36} &
\multicolumn{1}{c}{34} &
\multicolumn{1}{c}{4.0} 
\\
\multicolumn{1}{c}{M2} &
\multicolumn{1}{c}{41} &
\multicolumn{1}{c}{39} &
\multicolumn{1}{c}{4.6} &
\multicolumn{1}{c}{42} &
\multicolumn{1}{c}{39} &
\multicolumn{1}{c}{4.6} 
  \\
  \multicolumn{1}{c}{M3} &
\multicolumn{1}{c}{38} &
\multicolumn{1}{c}{36} &
\multicolumn{1}{c}{4.2} &
\multicolumn{1}{c}{41} &
\multicolumn{1}{c}{38} &
\multicolumn{1}{c}{4.5} 
  \\
  \multicolumn{1}{c}{M4} &
\multicolumn{1}{c}{47} &
\multicolumn{1}{c}{45} &
\multicolumn{1}{c}{5.2} &
\multicolumn{1}{c}{46} &
\multicolumn{1}{c}{43} &
\multicolumn{1}{c}{5.1} 
 \\
\multicolumn{1}{c}{M5} &
\multicolumn{1}{c}{50} &
\multicolumn{1}{c}{48} &
\multicolumn{1}{c}{5.5} &
\multicolumn{1}{c}{47} &
\multicolumn{1}{c}{46} &
\multicolumn{1}{c}{5.4} 
  \\
\hline
\hline
 
\end{tabular}
\end{center}
\end{table}

Next, I assume the fractional transit depth $A$ caused by the target feature  is given by:
\begin{equation}
    A(\lambda) \approx \frac{2 n_H(\lambda)H R_{\rm p}}{{R_{\rm s}}^2}
\end{equation}
where $n_H$ varies according the host star spectral subclass (for M0 we use the M1 value) and channel (Table \ref{table: spec feat}) and $H$ varies for each planet according to its $T_{\rm {eq}}$, $g$ and $\mu$.
I define the SNR for detection of the target feature in one transit observation, $SNR_1(\lambda)$, at any wavelength as:
\begin{equation}
    SNR_1 (\lambda) = \frac{A (\lambda)}{\sigma_A (\lambda)} 
\end{equation}
where $\sigma_A$ is the noise on the measurement of $A$. I estimate $\sigma_A$ as follows. 

I adopt a simplified representation of the the spectral feature, modelling it as `box car' of height $A(\lambda)$ and width $\Delta \lambda_A (\lambda)$ (where $\Delta \lambda_A$ = 0.5 FWHM, i.e. 0.15 \textmu m in Channel A and 0.6 \textmu m in Channel B), rising above a flat baseline. 
If sampled at the intrinsic resolving power of the instrument, $R$, the number of such samples across the spectral bin of width $\Delta \lambda_A$ is:
\begin{equation}
    n_A (\lambda) = \frac{R \Delta \lambda_A (\lambda)}{\lambda} 
\end{equation}
If the fractional transit depth (due to both atmosphere and planet) at a given wavelength is $\tau(\lambda)$, then the noise on this at the intrinsic resolving power of the instrument is $\sigma_{\tau}(\lambda)$. Assuming negligible uncertainty on the spectral baseline, the uncertainty on $A$ is then approximately:
\begin{equation}
    \sigma_A (\lambda) \approx \frac{ \sigma_{\tau} (\lambda)}{\sqrt{n_A (\lambda)} }
\end{equation}
To find $\sigma_{\tau}$ I assume that each transit observation consists of an in-transit period, $T_{\rm 14}$, and an out-of-transit period equal to $x T_{\rm 14}$.  As in the previous detection model, I assume a `box car' model of the transit, with 100\% duty cycle efficiency and only photon noise included. The photon-conversion efficiency, $\kappa$, is set to 0.48. The total collecting area is given by $A_{\rm {tel}} = \pi(D_{\rm {tel}}/2)^2$.  For these calculations $D_{\rm {tel}}$ was set to 30 m, and the SNR results subsequently scaled for different values of $D_{\rm {tel}}$ as described later.  The total number of photoelectrons per spectral resolution element per transit observation, $S (\lambda)$, is then given by:  
\begin{equation}
  S(\lambda) = \kappa A_{\rm {tel}} F_{\rm tel}(\lambda) \frac{\lambda}{hc}\frac{\lambda}{R} \left( [1-\tau(\lambda)]T14 +xT14 \right)
\end{equation}
where $F_{\rm {tel}} (\lambda)$ is the flux density received at the telescope calculated using a blackbody function, so that $F_{\rm {tel}} (\lambda) = \pi B_{\lambda}(T_{\rm {s}})(R_{\rm s}/d)^2$, and $\lambda/R$ is the width of the spectral resolution element. Since $\tau << 1$ we can simplify this to: 
\begin{equation}
  S(\lambda) \approx   \kappa A_{\rm {tel}} F_{\rm {tel}}(\lambda) \frac{\lambda}{hc}\frac{\lambda}{R}(1+x)T14
\end{equation}
The noise on the transit depth, $\sigma_\tau$, is then given by:
\begin{equation}
  \sigma_{\tau}(\lambda) = \frac{1}{\sqrt{S(\lambda)}} \frac{1+x}{\sqrt{x}}
\end{equation}
In these simulations I adopt $x$ = 4.  This gives a long baseline per transit to reduce photon noise (assuming a very low noise floor), and thus the uncertainty on the transit depth, and also for the characterisation and correction of any stellar variability.

For each planet the minimum $SNR_1$ across both channels is obtained, ${SNR_1}_{min}$, and used to calculate the number of transits, $N_{\rm t}$. Assuming that $\sigma_{\tau}(\lambda)$ falls with the square root of the number of transit observations and a 3$\sigma$ detection threshold [as used in \cite{Rauer2011}]:
\begin{equation}
    N_{\rm t} = \left(\frac{3}{{SNR_1}_{min}}\right)^2
\end{equation}
Thus when this number of transit observations have been co-added for a given planet, the target spectral feature will be detectable at an SNR of 3 or more in every spectral bin in boths channels.
Once $N_{\rm t}$ has been calculated for each transiting planet we proceed to a scheduling algorithm.

\subsubsection{Scheduling}
 
While each planet needs to complete its required number of transit observations ($N_{\rm t}$) the scheduling of these must take into account the transit ephemerides of other planets in the sample. The schedule must avoid any overlaps in observation times between different planets and must minimize the the total time taken to complete all required transit observations in the sample.

The scheduling is performed for two different conditions. The first considers an `unrestricted' FOR so that all planets in the sample are observable at all times.  This would approximate the situation for a lunar polar telescope or novel deep-space observatory design at L2 operating in the shadow of a detached Sun-shield as described previously.  The second considers a `restricted' FOR and applies for an observatory at L2 with an attached Sun-shield and an architecture that permits observations from 90-180$^\circ$ in solar elongation, giving 50\% sky instantaneous sky coverage, centred on the anti-Sun axis.  This compares to 85-135$^\circ$ solar elongation for JWST, which gives it 39\% instantaneous sky coverage.  The direction of the centre of the FOR changes with time and has a period of 365.25 days.  As the observatory orbits the Sun, different planets in the sample will be inside or outside the the FOR.

For each realisation the scheduling algorithm functions as follows.  For a given planet, the total duration of one transit observation equals $(1+x) T_{14}$. For each planet, the start time for the first transit observation is chosen randomly from within its period, the first central transit time being $0.5(1+x)T14$ longer than this. Each planet is also randomly assigned an ecliptic longitude. For the restricted FOR case, at any time, the observatory has an ecliptic longitude, $\theta$, i.e. the direction of the centre of the FOR.   Starting with $\theta$ = 0$^\circ$, the rate of change of $\theta$ is 360/365.25 degrees per day.
Only planets with ecliptic longitudes within $\pm$90$^\circ$ of $\theta$ will be viewable at any time.  

The algorithm proceeds as follows for the unrestricted FOR case.
An initial timeline is constructed of length = 5 $\times$ the maximum value of \Nt $\times$ $P$ in the sample. 
For each planet, all its potential transit observations are initially marked on this timeline (start and end times of each transit observation).    
The planet with the longest period is then chosen as the first `reference' planet.  The reference planet transits are truncated at \Nt transits (i.e. all transits > \Nt are removed from the timeline).  The reference planet's remaining transit observation times are compared with each of the other planets in order of decreasing period.  Any overlapping transit observations are deleted from the timelines of the shorter period planets, such that the end of this process the reference planet should have no observing clashes with any of the other planets.  The planet with the next longest period now becomes the reference planet and the process is repeated, truncating its transits to \Nt, comparing its timeline to those of shorter period planets and removing clashes. This process is repeated with the reference planet chosen in order of period until the shortest period planet becomes the reference planet.  At that stage there should be no clashing observations and all planets will have \Nt transit observations in their timelines.

For the restricted FOR case the same algorithm is used with the following modification. The periods where the planet will be out of the instantaneous FOR are marked on the timeline for each planet and any transit observations that are in or cross into the out-of-view periods are deleted from the timeline.  The algorithm then proceeds as for the unrestricted FOR case, so that the final schedule has the required number of transit observations for each planet, \Nt, there are no clashes between planets, and all scheduled observations will be within the instantaneous FOR of the observatory.

Once the algorithm is completed, the mission is fully scheduled, and the mission duration, \Tmiss, is given by the time when all planets have completed their transit observations.  The total mission time does not include any time for commissioning, and it is assumed that slewing and housekeeping functions are accommodated in the time between active observations.

\subsubsection{Type 1 and type 2 simulations}

We perform scheduling and obtain \Tmiss results independently for the cases where $\mu$ is fixed (type 1) and for where $\mu$ is random (type 2).  In the calculation of \Nt (and thus the scheduling and calculation of mission duration) $\mu$ is the only factor that cannot be reasonably estimated  a priori.   Hence the type 1 case reflects a mission scheduled without a priori knowledge of $\mu$.   The type 2 simulations produce a schedule based on the actual $\mu$ of each planet and is thus closer to the `ground truth'.
 Despite this, the type 2 schedule would not be easy to produce in reality (i.e. in planning the actual mission) relying as it does on knowing the actual $\mu$ for each planet.  We could envisage a situation however where, starting with a schedule that has an indefinite number of transits per planet, as the mission proceeds the atmospheric SNR for each planet is determined experimentally, and when the goal SNR is reached it is deemed to have completed \Nt transits and its remaining transits removed from the schedule. The schedule is then recalculated to account for the removed observations. \Nt is effectively found `experimentally' as the spectra are analysed.  The final schedule thus obtained should then approximate that simulated here.  We can compare the results for both type 1 and type 2 schedules in this study to see if the more practical type 1 schedule gives results close to the more realistic type 2 schedule.

\subsubsection{\Npool}
\begin{figure}
 \begin{center}
 	\includegraphics[trim={0cm 0cm 0cm 0cm}, clip, width=1.0\columnwidth]{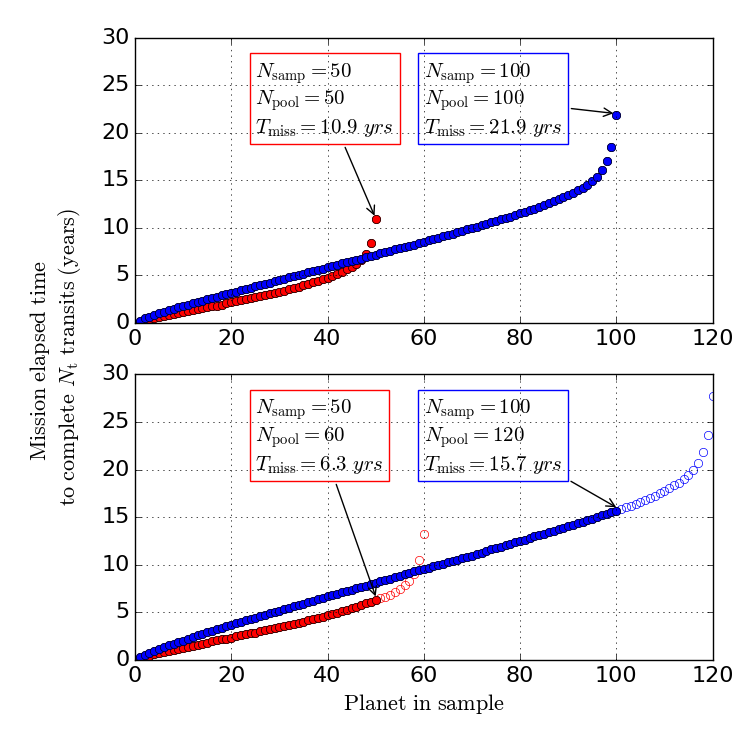}
 \end{center}
    \caption{The `uptick' effect on the mission elapsed time. Results shown for \Dtel = 20 m with a type 2 simulation and an unrestricted FOR.
    Dots give the average mission elapsed time for the $n$th planet in the time-ordered sequence of a scheduled sample to complete its \Nt transits.  
    A small number of planets at the end of the ordered sequence take up a disproportionate amount of mission time. Upper plot: The effect shown for \Nsamp = 50 and \Nsamp = 100.  Lower plot: Mitigation of this effect by including \Npool planets in the initial schedule but completing mission when \Nsamp is achieved, reducing mission duration. In this case  \Npool = 1.2 $\times$ \Nsamp. }
\label{fig:pool}
\end{figure}

It was found that simply scheduling a sample of \Nsamp planets resulted in the small number planets of the sample pushing up the mission duration disproportionately. This effect is shown in Figure \ref{fig:pool} (upper plot) for \Nsamp = 50 and \Nsamp = 100 using \Dtel = 20 m with an unrestricted FOR and type 2 simulations. In each of the 100 realisations of the final schedule, the planets are ordered by the mission elapsed time at which they complete their required number of transits, \Nt. The mean time for the $n$th planet in the sequence is then obtained and plotted as shown in Figure \ref{fig:pool} (upper plot).  The time for the final planet gives the mission duration, \Tmiss. 
Initially the mission elapsed time increases with planet in a near linear way, however for the final few planets in the sequence the mission elapsed times are disproportionately higher leading to an `uptick' in the curve that pushes up the final \Tmiss.  This effect occurs also in the restricted FOR schedule simulations.

Considering the unrestricted FOR case, the minimum time to complete all of a planets observations is \Nt $\times$ $P$, although this may be increased by the scheduling algorithm if there are clashing transit observations with other planets. In any sample there is a range of \Nt $\times$ $P$. It is the small number of high \Nt $\times$ $P$ planets (toward the end of the sequences shown) that tend to push up the mission time disproportionately in the context of the scheduling algorithm.

If the sample size is expanded, we bring in some new planets which have lower \Nt $\times$ $P$ than the highest \Nt $\times$ $P$ planets in the smaller sample.  Therefore in the expanded sample the $n$th planet may be completed sooner than than in the smaller sample, since the $n$th planet has a lower \Nt $\times$ $P$ in the expanded sample.  This is particularly noticeable when $n$ is close to \Nsamp of the smaller sample.  This can be seen in \ref{fig:pool} (upper plot) where the 50th planet is completed earlier in the \Nsamp = 100 sample than the \Nsamp = 50 sample.  This `uptick' effect thus has the potential to significantly increase \Tmiss to the detrimental of overall mission feasibility.

To mitigate this, we can instead initially schedule more planets than \Nsamp itself and cut off the mission when \Nsamp planets have completed their observations.  This way the best performing planets are selected from a slightly larger `pool' of planets of size \Npool.  In Figure \ref{fig:pool} (lower plot) we used  \Npool = 1.2 $\times$ \Nsamp. For example, in a mission with \Nsamp = 50, we initially schedule \Npool = 60 planets.  We can see that \Nsamp planets are completed just before the `uptick' effect kicks in, largely mitigating it.  This significantly improves \Tmiss (at the expense of requiring more planets to be known at the start).  Given this benefit to mission times, the final scheduling was performed with \Npool = 1.2 $\times$ \Nsamp in all cases, and \Tmiss obtained when the $n$th ($n$ = \Nsamp) planet in the sequence completes all its \Nt observations.

\subsubsection{Varying \Dtel}
To investigate how  $T_{\rm {miss}}$  varies with telescope aperture for different \Nsamp values, I take the 
$SNR_1$ values obtained using the simulation with \Dtel = 30 m, for each of the 100 realisations, and scale these to different \Dtel values (\Dtel = 10, 15, 20, 30, 40 and 50 m) by multiplying the results for the baseline case by $D_{\rm {tel}}/30$.  ${SNR_1}_{min}$ and $N_{\rm t}$ are obtained for each case, the planets ordered by detectability and then run through the scheduling algorithm for different \Nsamp values as above.  For each \Dtel and \Nsamp combination, 100 values for $T_{\rm {miss}}$ are thus obtained and the mean and standard deviation found for each case.

\begin{figure}
 \begin{center}
 	\includegraphics[trim={0cm 0cm 0cm 0cm}, clip, width=1.0\columnwidth]{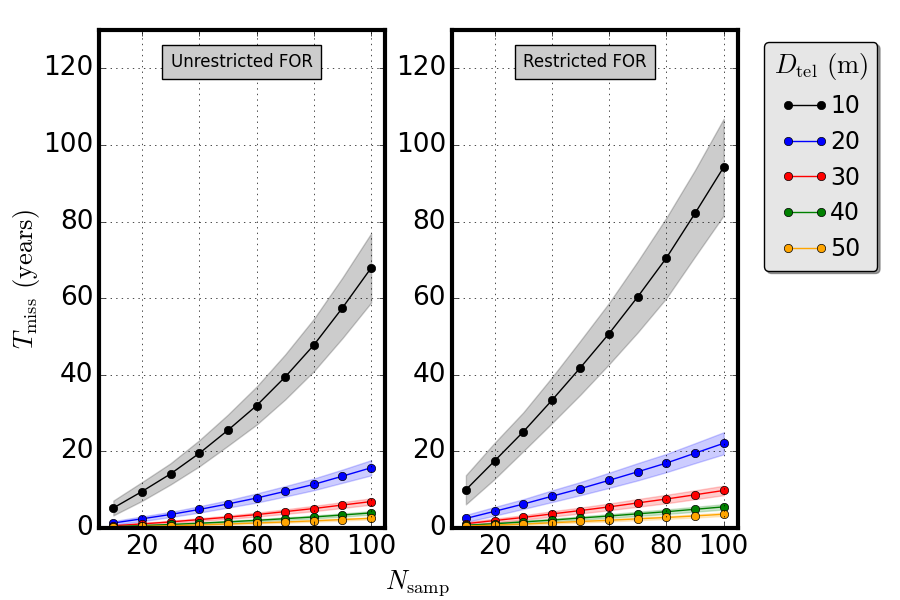}
    {\hspace{0.1cm}A: Type 1 result   }
 	\includegraphics[trim={0cm 0cm 0cm 0cm}, clip, width=1.0\columnwidth]{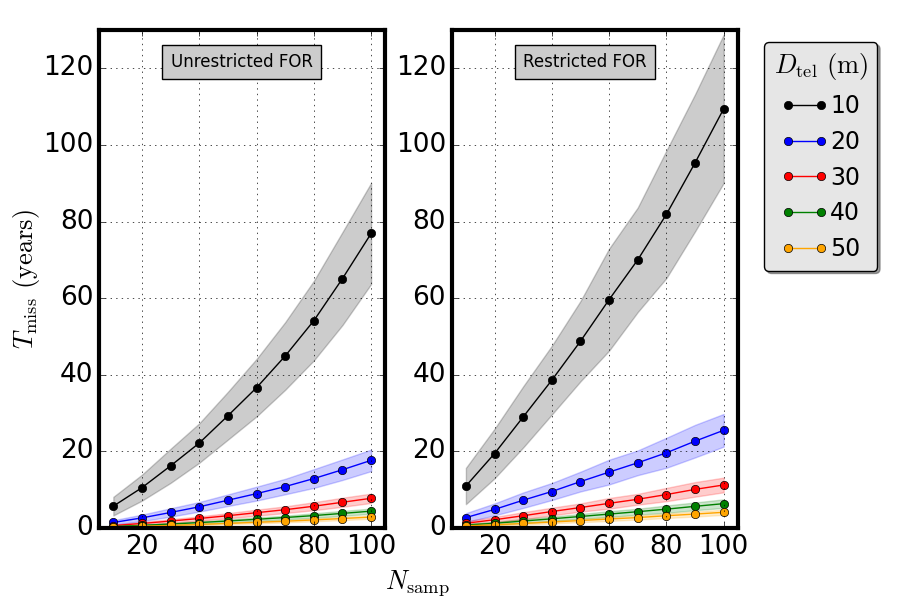}
 	{\hspace{0.1cm}B: Type 2 result   }
 \end{center}
    \caption{Mission duration, \Tmiss, vs sample size, \Nsamp, for different primary mirror sizes, \Dtel. Dots show mean results obtained over 100 realizations, with 1$\sigma$ range shown as shaded area.  A: Type 1 simulations. B: Type 2  simulations. Left-sided plots: Unrestricted field-of-regard. Right-sided plots: restricted field-of-regard.}
\label{fig:T_miss}
\end{figure}  

\begin{figure}
 \begin{center}
 	\includegraphics[trim={0cm 0cm 0cm 0cm}, clip, width=1.0\columnwidth]{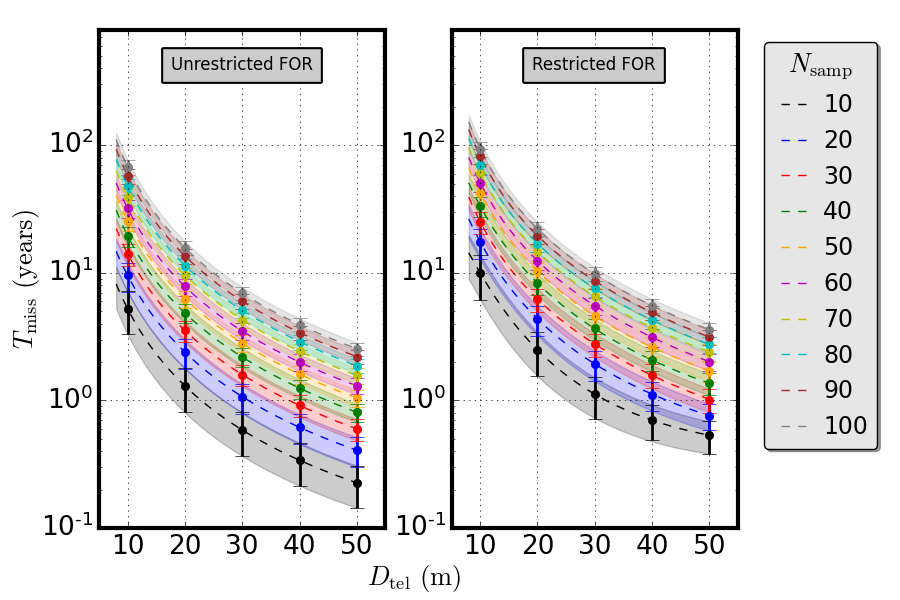}
 	    {\hspace{0.1cm}A: Type 1 result   }
 	\includegraphics[trim={0cm 0cm 0cm 0cm}, clip, width=1.0\columnwidth]{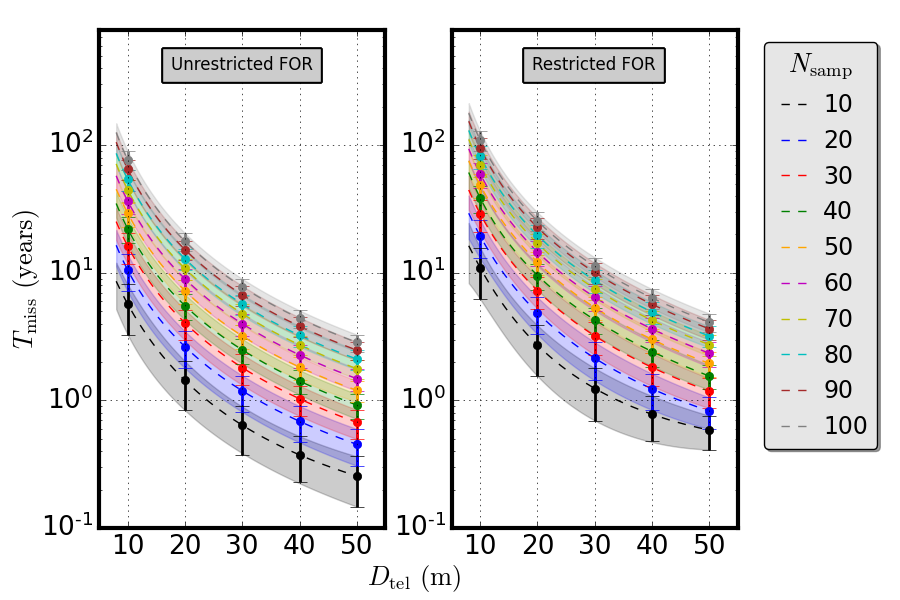}
 	    {\hspace{0.1cm}B: Type 2 result   }
 \end{center}
    \caption{Mission duration, \Tmiss, vs primary mirror size, \Dtel, for different sample sizes, \Nsamp.  Dots show the  mean result over 100 realisations and error bars show 1$\sigma$ range.  Polynomial fits to the mean values are shown with the dotted lines.  Polynomial fits to the maximum and minimum error bar values are shown by the edges of the shaded regions.  
   A: Type 1 simulations. B: Type 2 simulations. Left-sided plots: Unrestricted field-of-regard. Right-sided plots: restricted field-of-regard. }
\label{fig:T_miss_vs_D_tel}
\end{figure}  

\begin{figure*}
 \begin{center}
 	\includegraphics[trim={0.5cm 0cm -0.5cm 0cm}, clip, width=1.5\columnwidth]{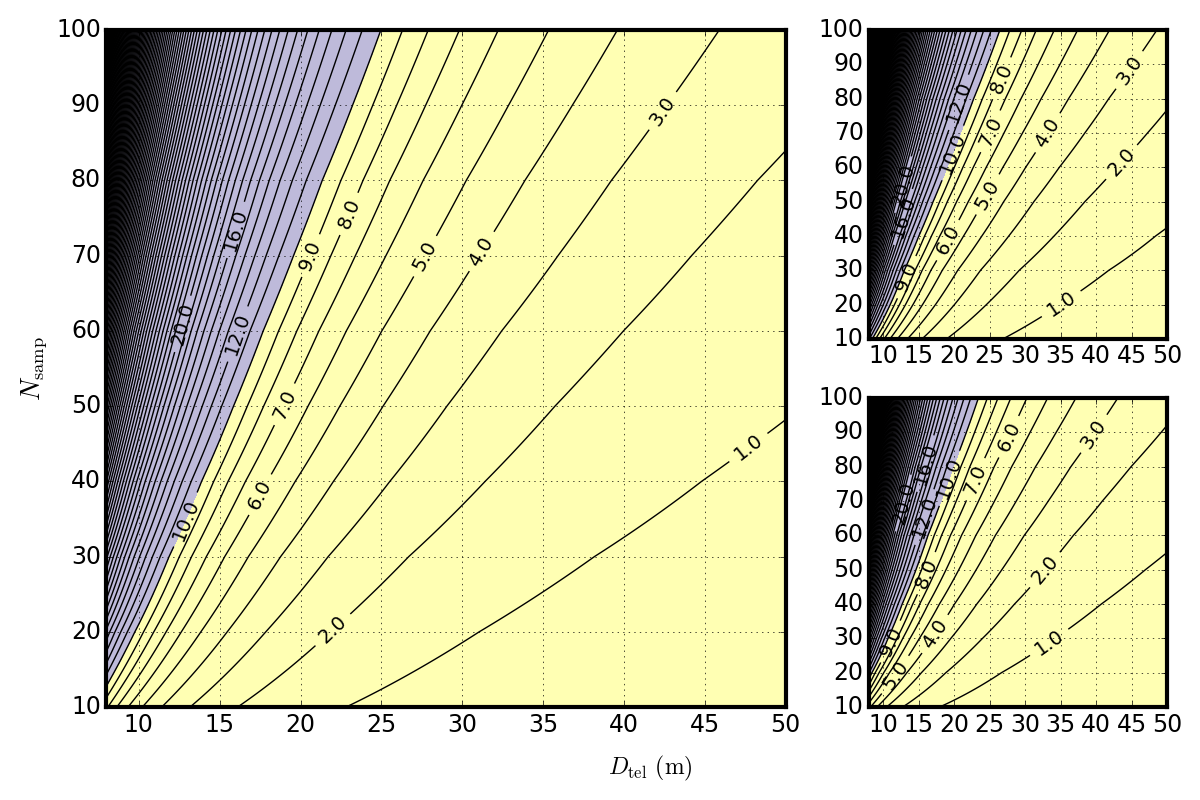}
 	    {\hspace{30cm}A: Type 1 result   }
	    
 	\includegraphics[trim={0.5cm 0cm -0.5cm 0cm}, clip, width=1.5\columnwidth]{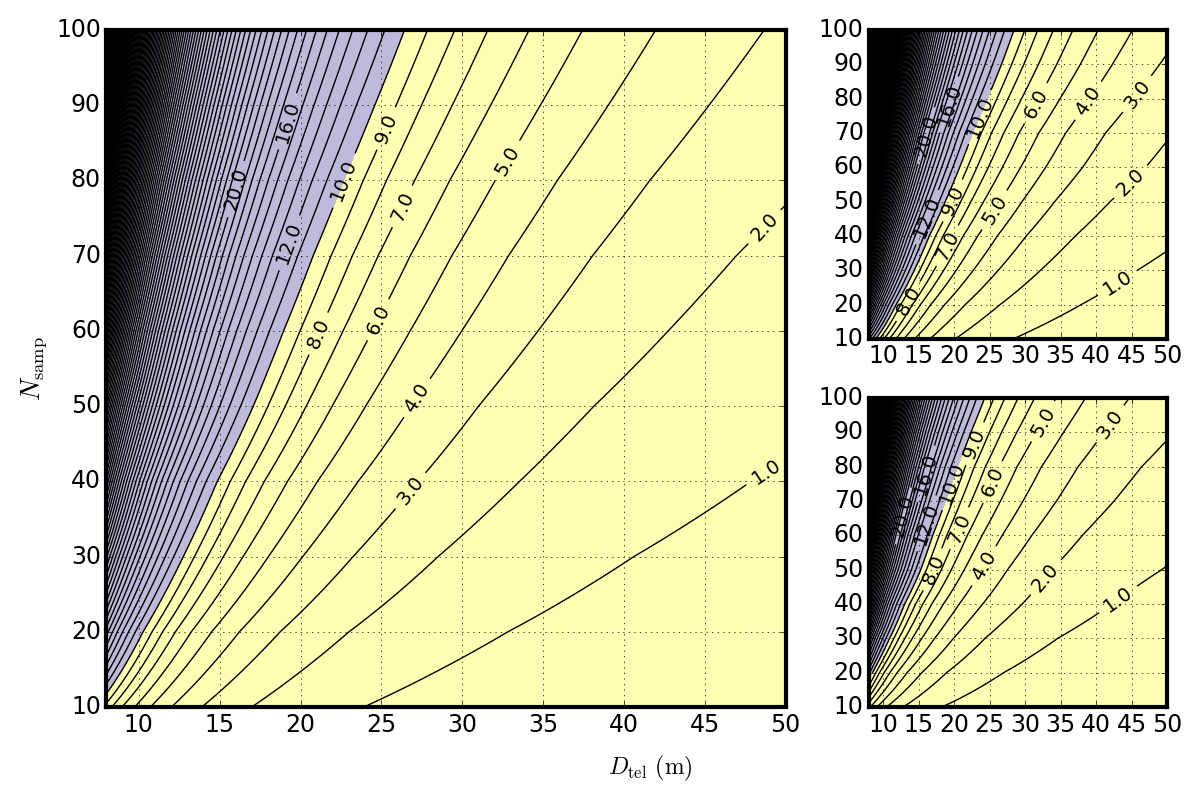}
 	    {\hspace{30cm}  B: Type 2 result   }
 \end{center}
    \caption{Contour plots of mission duration, $T_{\rm {miss}}$, (in years) vs primary mirror size, \Dtel, and sample size, \Nsamp, for the unrestricted field-of-regard case. A: Type 1 simulation results. B: Type 2 simulation results. 
    The yellow-shaded region indicates a mission duration < 10 years and the purple-shaded region indicates a mission duration > 10 years.  Left-sided plots: mean result.  Smaller upper right plots: upper 1$\sigma$ limit. Smaller lower right plots: lower 1$\sigma$ limit.   }
\label{fig: cplot1}
\end{figure*}

\begin{figure*}
 \begin{center}
    \includegraphics[trim={0.5cm 0cm -0.50cm 0cm}, clip,width=1.5\columnwidth]{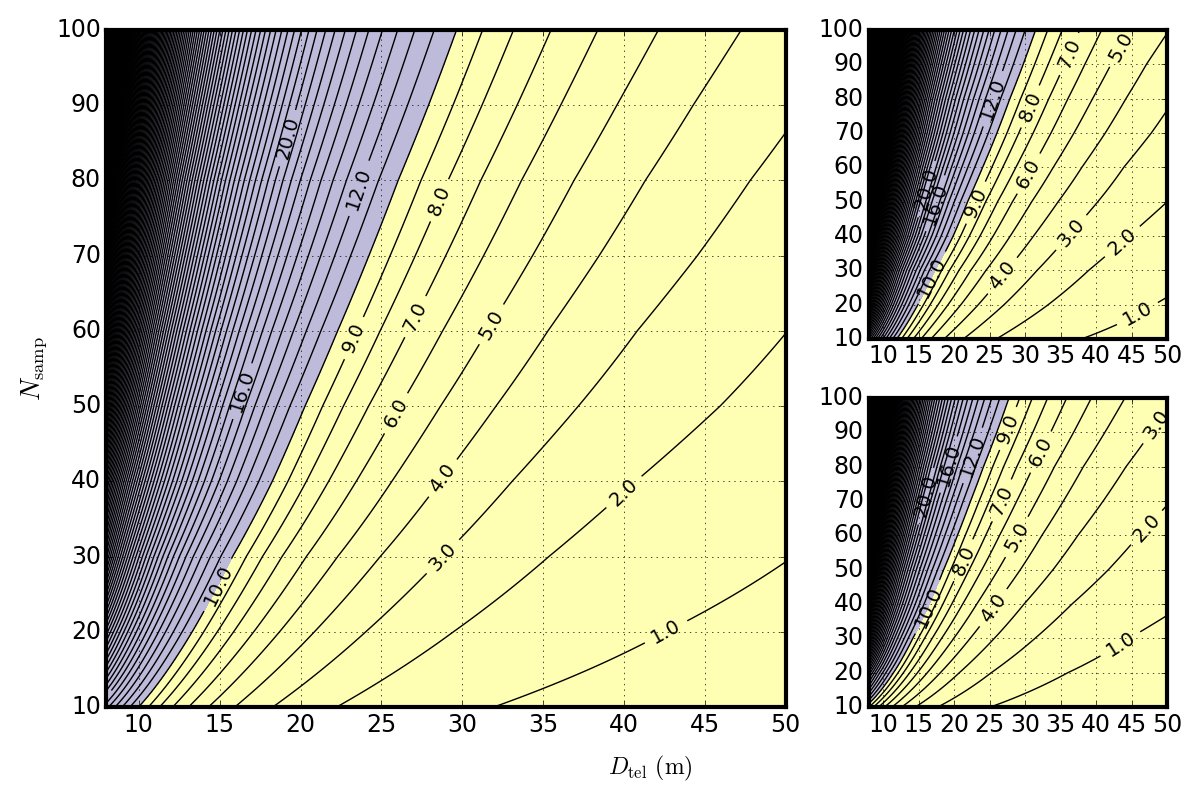}
 	    {\hspace{30cm}A: Type 1 result   }
 	    
    \includegraphics[trim={0.5cm 0cm -0.50cm 0cm}, clip,width=1.5\columnwidth]{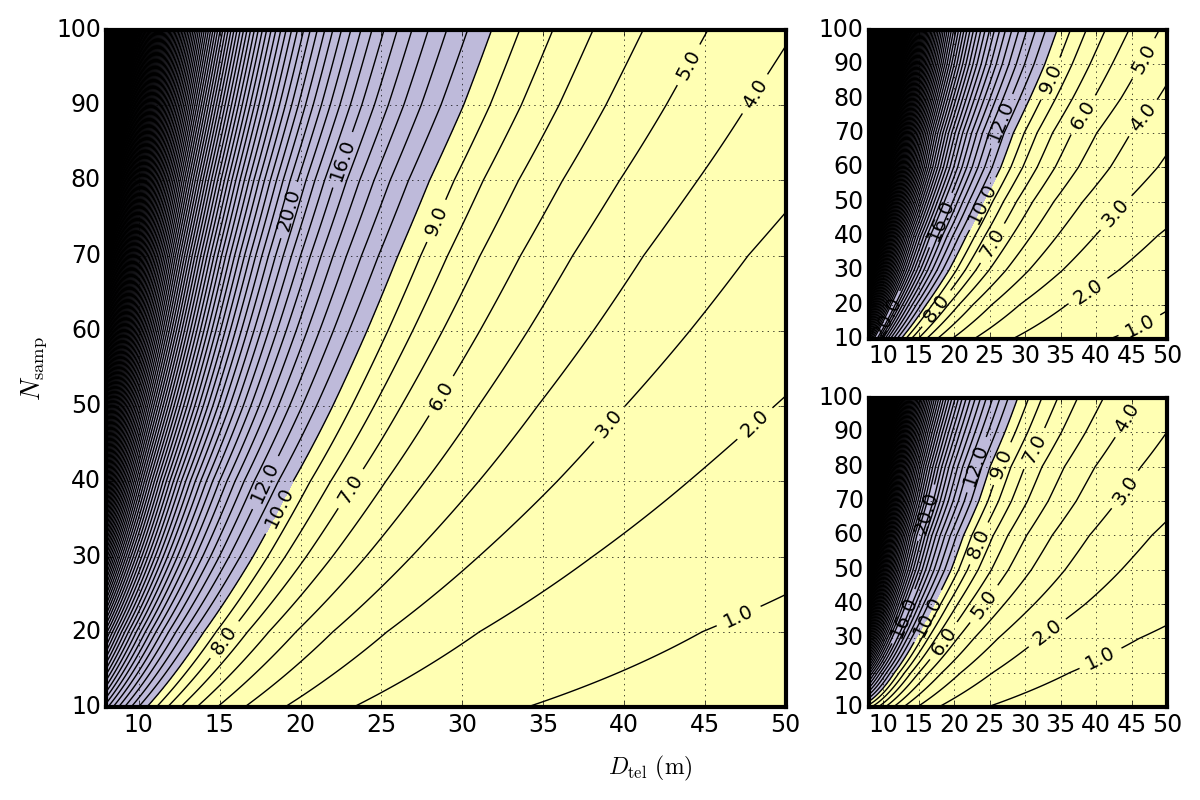}
 	    {\hspace{30cm}  B: Type 2 result   }
 \end{center}
    \caption{Contour plots of mission duration, $T_{\rm {miss}}$, (in years) vs primary mirror size, \Dtel, and sample size, \Nsamp, for the restricted field-of-regard case. A: Type 1 simulation results. B: Type 2 simulation results. 
    The yellow-shaded region indicates a mission duration < 10 years and the purple-shaded region indicates a mission duration > 10 years.  Left-sided plots: mean result.  Smaller upper right plots: upper 1$\sigma$ limit. Smaller lower right plots: lower 1$\sigma$ limit.   }
\label{fig: cplot2}
\end{figure*}

\subsection{Results}

Figure \ref{fig:T_miss} shows  \Tmiss vs \Nsamp for different \Dtel  values. The mean results over 100 realisations and 1$\sigma$ confidence region are displayed in each case.  Results are shown for 
 fixed $\mu$ (type 1) and randomised $\mu$ (type 2) simulations, for both the restricted and unrestricted FOR cases. We can see that type 1 results give slightly shorter \Tmiss values than type 2. As may be expected the restricted FOR results in much longer \Tmiss than the corresponding unrestricted FOR case.

In Figure \ref{fig:T_miss_vs_D_tel} these $T_{\rm {miss}}$ results are plotted vs \Dtel
for different sample sizes, \Nsamp.  
The dots and error bars give the mean and 1$\sigma$ range respectively from 100 realisations.
3rd order polynomials are fitted in log-log space to the mean values and also to the maximum and minimum values of each error bar. The resulting equations 
allow us to infer the mean and 1$\sigma$ range of \Tmiss for \Dtel values between and beyond those  obtained directly through the simulations.  We use these to obtain average $T_{\rm {miss}}$ values over a finer \Dtel grid (with a gradient of 0.1 m) giving the dotted lines shown. The corresponding 1$\sigma$ ranges are shown by the shaded areas.   The mean and 1$\sigma$ range lines are also extended to smaller \Dtel values down to 8 m as shown. These lines using the finer \Dtel grid are then  used to generate the contour plots\footnote{The matplotlib `contour' function is used.} in Figures \ref{fig: cplot1} (unrestricted FOR) and \ref{fig: cplot2} (restricted FOR).  These allow $T_{\rm {miss}}$ to be estimated for any combination of \Nsamp and \Dtel. The regions for $T_{\rm {miss}}$ < 10 years and > 10 years are demarcated.  I adopt 10 years as maximum viable mission duration for a mission at L2 (assuming no maintenance of the observatory).

Results for selected cases are summarised  in Table \ref{table: summary}, combined with CI estimates from Section \ref{results 4.2}.  Four observatory types are considered: 1) an L2 observatory with restricted FOR, 2) an L2 observatory with unrestricted FOR, 3) a lunar observatory at one pole, 4) two lunar observatories, one at each pole.  The results obtained here for the restricted FOR and unrestricted FOR are directly applicable to the first and second L2 types respectively.
For two lunar telescopes we can also use the unrestricted FOR results.
In Table \ref{table: summary} the minimum number of planets needed, $N_{\rm min}$, takes into account the fact that both the type 1 and type 2 schedules are based on selecting the best performing planets from a pool of planets, \Npool, of size 1.2 $\times$ \Nsamp. In most of these cases $N_{\rm min}$ = \Npool.
For a single lunar telescope we can use the unrestricted FOR results with the caveat that since only half the sky will be observable $N_{\rm min}$ = 2 $\times$ \Npool assuming an even distribution over the whole sky. 
 
These results of course reflect the instrument design chosen, the various assumptions made for the the modeling, the metrics for detection (including the need to achieve adequate SNR out to 11 \textmu m), and the efficiency of scheduling. There is quite a bit of variability around the mean cases as shown by the right-hand plots in Figures \ref{fig: cplot1} and \ref{fig: cplot2}.  Thus depending on the exact details of the actual planet population that will be available at the time of the mission, the final mission times may be greater or less than the predicted mean results.

\begin{figure}
 \begin{center}
 	\includegraphics[trim={0cm 0cm 0cm 0cm}, clip, width=1.0\columnwidth]{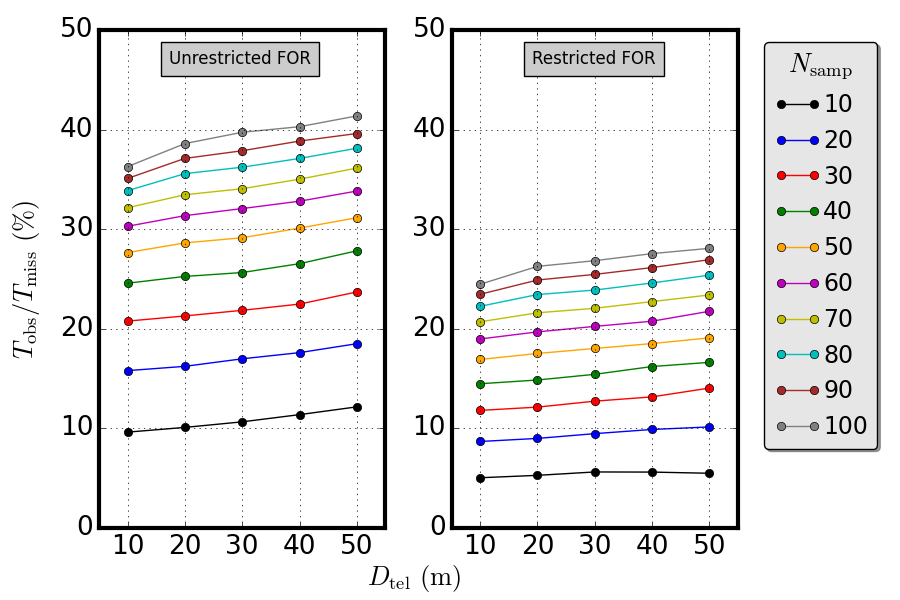}

    {\hspace{0.1cm}A: Type 1 result   }
 	
 	 \includegraphics[trim={0cm 0cm 0cm 0cm}, clip, width=1.0\columnwidth]{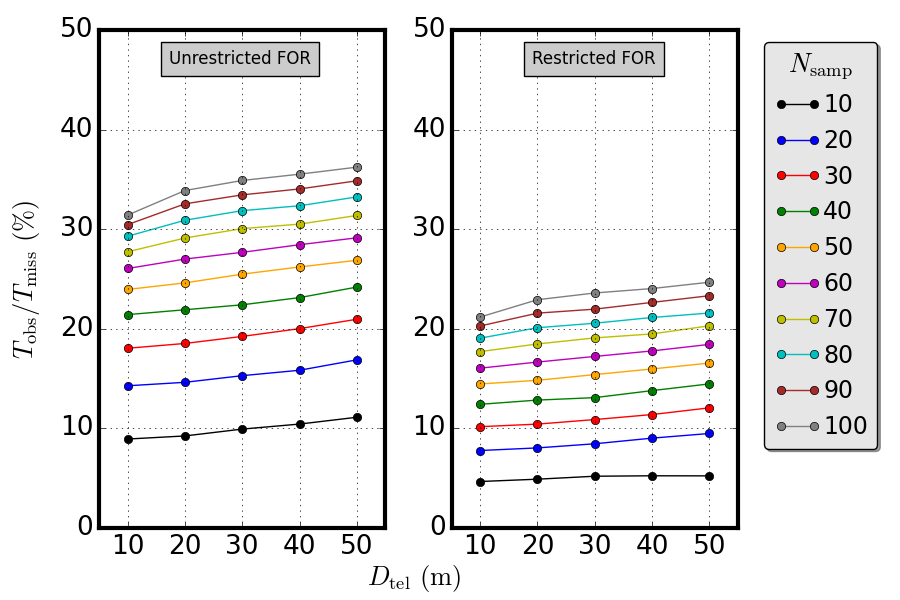}
 	 
    {\hspace{0.1cm}B: Type 2 result   }

 \end{center}
    \caption{Ratio of total observing time, \Tobs, to total mission time, \Tmiss, vs primary mirror size, \Dtel, for different sample sizes, \Nsamp. Dots give the mean result over 100 realisations.
    A: Type 1 simulations. B: Type 2  simulations. Left-sided plots: Unrestricted field-of-regard. Right-sided plots: restricted field-of-regard.}
\label{fig: ratio}
\end{figure}  

\subsubsection{10-m telescope}
For the unrestricted FOR, if we look at the lower 1$\sigma$ limit results (the optimistic limit in terms of mission duration) we find that it is unlikely that any combination of \Nsamp > 27 and \Dtel < 10 m will result in a mission of < 10 years duration using the type 1 results (Figure \ref{fig: cplot1} A lower right). On average a 10-m telescope delivers a 19 (type 1 result) to 21 (type 2 result) planet survey in 10 years (Figure \ref{fig: cplot1} A left and B left).  

For the restricted FOR, with increased mission times, the yields are lower for the same duration.  Taking the most optimistic lower 1$\sigma$ limit, both the type 1 and type 2 results indicate that it is unlikely that any combination of of \Nsamp > 16 and \Dtel < 10 m will result in a mission of < 10 years duration (Figure \ref{fig: cplot2} A lower right and B lower right).  The average results indicate that a 10-m telescope would deliver a 10-planet survey in 10 years going by the type 1 result (Figure \ref{fig: cplot2} A left) or an 8-planet survey by the type 2 result.  The latter value was obtained by fitting a 2nd order polynomial to points along the 10-year demarcation line (Figure \ref{fig: cplot2} B left) and using this to predict the result falling just outside the boundaries of the chart.

 \subsubsection{20-planet survey}
Considering \Nsamp = 20, we can see from Table \ref{table: summary} that 15-m class telescopes are very likely to be able to complete a 20-planet survey within 10 years. Taking the worst case result, the L2 observatory with restricted FOR and type 2 simulation, we find that a 10-year mission is achieved with a 14.0 (+2.1/-2.4) m telescope.  15-m telescopes could complete the 20-planet survey in anything from 4.3 ($\pm$1.1) years to 8.7 ($\pm$2.7) years depending on the observatory type.  10-m class telescopes might be able to achieve a 20-planet survey in 10 years in the best case scenario, using the unrestricted FOR L2 observatory or the lunar telescope cases.  A survey of this size could constrain the \fL 95\% CI size from 0.13 to 0.4 depending on the observed value.

\subsubsection{15-m telescope}
If we consider what a 15-m telescope can achieve in a 10 year mission, we find that using the slightly more pessimistic type 2 results it can complete an \Nsamp of 23 (+8/-6) planets for the restricted FOR L2 observatory, increasing to 40 (+9/-7) for the unrestricted FOR cases.  If we consider an \fLo of 0.05, i.e. one planet in the survey appears positive for life, the 23 planet survey result would constrain the upper limit of \fL to to $\sim$ 0.19 at 95\% confidence, while 40 planets would constrain this to $\sim$ 0.15. The lower bounds in each of the previous cases are <0.01. Thus there is only a modest improvement in \fL precision between the two cases, so that  a 15-m telescope in a `traditional' L2 configuration with restricted FOR performs close to one with unrestricted FOR in this regard.

\subsubsection{50-planet survey}
However, when we increase the sample size to 50, 15-m class telescopes are much less likely to succeed within 10 years. Table \ref{table: summary} shows that \Dtel would need to be 15.9 m (+1.1/-1.3) in the best case scenario, type 1 with an unrestricted FOR.  If we go by the more pessimistic type 2 results of \Dtel = 17.1(+1.6/-1.8) m, we can say that an \Nsamp of 50 probably calls for at least a 17-m telescope. Similarly looking at the type 2 result for the restricted FOR case, \Dtel = 22.0 (+2.1/-2.4) m, a 22-m telescope is probably required for \Nsamp = 50 in this situation. 
A 50-planet survey could constrain the upper limit of \fL to $\sim$ 0.06 at 95\% confidence if the sample was completely negative, or between $\sim$  0.03-0.2 if \fLo was 0.1 (i.e. 5 positive planets in the sample).

\subsubsection{100-planet survey}
Finally, Table \ref{table: summary} summarises the results for a 100-planet survey.  This sample size demands \Dtel
of between $\sim$ 25 to 32 m for 10 year missions depending on the simulation type and the field-of-regard.  However, given the predictions of TESS yields \citep{Barclay2018, Sullivan2015}, it is possible that very large sample sizes ($>>$ 50) may simply not exist in the coming decade.  Also, as shown earlier, the incremental improvement on the CI of \fLo falls with sample size.  Table \ref{table: summary} shows only a modest reduction of CI size for \Nsamp = 100 compared to \Nsamp = 50.  For these reasons planning a mission around a 50-planet survey may be optimal.

\subsubsection{Total observing time} 
The total observing time, \Tobs, is the sum of all the time spent observing all of the planets in the sample.  For each planet this equates to $(1+x)T_{14} \times$ \Nt. 

In Figure \ref{fig: ratio} we plot the ratio of mean $T_{\rm {obs}}$ over 100 realisations  to the
mean $T_{\rm {miss}}$ over 100 realisations  vs \Dtel for different \Nsamp values.  This serves as a measure of the efficiency of the mission in terms of observing time.  The unrestricted FOR case is more efficient since \Tmiss values are smaller compared to the restricted FOR.  The type 2 simulations return slightly less efficient missions than type 1 which may be related to the longer \Tmiss times for the type 2 simulations.  Efficiency improves with \Nsamp and generally with larger \Dtel (although this trend is less apparent in the restricted FOR cases for \Nsamp = 10).  A wide range of efficiency is seen ranging from $\sim$ 5\% (for \Nsamp = 10 in the type 2 restricted FOR case) to $\sim$ 41\% (for \Nsamp = 100 in the type 1 unrestricted FOR case with a \Dtel of 50 m). If we consider a mission with a goal \Nsamp of 50, then if we use a 17-m telescope with an unrestricted FOR, e.g. the novel L2 design or the lunar telescope cases, the efficiency is around 28\% from type 1 simulations or 24\% from type 2 simulations.  If we consider a more traditional L2 setup with a restricted FOR then using a 22-m telescope the efficiency for \Nsamp = 50 is 18\% from the type 1 simulation or 15\% from the type 2 simulation.  Also if the amount of OOT time per observation $xT_{14}$  (where $x$ = 4 for these simulation)  is reduced then the efficiency will drop further.  While some of the off-target time will be needed for housekeeping, calibration and slewing time, these ratios of \Tobs to \Tmiss indicate a low efficency in terms of utlising the full mission time.
This is an inevitable limitation of using the transit spectroscopy approach since long stretches of time may elapse while waiting for the next available transit, and the active observing time is a small fraction of a planet's period.  The time between observations could of course be used for alternative science goals.  Therefore these results indicate that the DRAKE mission should probably be an integral part of a more wide-ranging observatory similar to NASA-sponsored Decadal Survey Mission Concept Studies such as the Origins Space Telescope \citep{Battersby2018}. It would also be consistent with the themes of ESA Voyage 2050 \citep{Favata2021} and could be a science driver for a new space- or Moon-based observatory within that programme.

\begin{table*}
\begin{center}
\caption{Summary of findings for selected cases. For each \Nsamp the corresponding 95\% CI is given for three different observed \fL values. The minimum number of known candidate planets, $N_{\rm min}$, needed is given for each observatory type.  
The minimum telescope primary mirror diameter, \Dtel, for a 10-year mission is given for each case, as well as mission duration, \Tmiss, for 15-m, 22-m and 30-m telescopes. Bracketed values give the 1$\sigma$ range.}
\label{table: summary}
\setlength{\tabcolsep}{5pt}
\begin{tabular}{lccccccccc} 
\hline
\hline
\multicolumn{1}{c}{\Nsamp} &
\multicolumn{3}{c}{95\% CI } &
\multicolumn{1}{c}{Observatory}
&
\multicolumn{1}{c}{$N_{\rm min}$}
&
\multicolumn{1}{c}{\Dtel}
&
\multicolumn{1}{c}{$T_{\rm miss}$}
&
\multicolumn{1}{c}{$T_{\rm miss}$}
&
\multicolumn{1}{c}{$T_{\rm miss}$}
\\

\multicolumn{1}{c}{ } &
\multicolumn{1}{c}{\fLo } &
\multicolumn{1}{c}{\fLo } &
\multicolumn{1}{c}{\fLo } &
\multicolumn{1}{c}{type}
&
\multicolumn{1}{c}{ }
&
\multicolumn{1}{c}{(10 yr) }
&
\multicolumn{1}{c}{(15-m) }
&
\multicolumn{1}{c}{(22-m) }
&
\multicolumn{1}{c}{(30-m) }
\\

\multicolumn{1}{c}{} &
\multicolumn{1}{c}{0.0 } &
\multicolumn{1}{c}{0.1 } &
\multicolumn{1}{c}{0.5 } &
\multicolumn{1}{c}{} &
\multicolumn{1}{c}{} &
\multicolumn{1}{c}{(m)} &
\multicolumn{1}{c}{(yrs)} &
\multicolumn{1}{c}{(yrs)} &
\multicolumn{1}{c}{(yrs)} 
\\

\multicolumn{1}{c}{} &
\multicolumn{1}{c}{} &
\multicolumn{1}{c}{} &
\multicolumn{1}{c}{} &
\multicolumn{1}{c}{} &
\multicolumn{1}{c}{} &
\multicolumn{1}{c}{Type 1} &
\multicolumn{1}{c}{Type 1} &
\multicolumn{1}{c}{Type 1} &
\multicolumn{1}{c}{Type 1} 
\\

\multicolumn{1}{c}{} &
\multicolumn{1}{c}{} &
\multicolumn{1}{c}{} &
\multicolumn{1}{c}{} &
\multicolumn{1}{c}{} &
\multicolumn{1}{c}{} &
\multicolumn{1}{c}{Type 2} &
\multicolumn{1}{c}{Type 2} &
\multicolumn{1}{c}{Type 2} &
\multicolumn{1}{c}{Type 2} 
  \\
\hline
\hline
20  &	0-0.13 & 0.02-0.28  & 0.3-0.7 &  L2 (Unrestricted FOR) & 24 &	9.8 (+1.2/-1.4) & 4.3$(\pm$1.1) & 2.0($\pm$0.5) & 1.1($\pm$0.3)  \\
 &	  &    &   &    &   &	10.3 (+1.5/-1.8) & 4.7$(\pm$1.5) & 2.2($\pm$0.7) & 1.2($\pm$0.4)  \\
 
& & & & L2 (Restricted FOR) & 24 &	13.3 (+1.6/-1.9) & 7.8$(\pm$2.0) & 3.6($\pm$0.9) & 1.9($\pm$0.5)  \\
&	  &    &   &    &   &	14.0 (+2.1/-2.4) & 8.7$(\pm$2.7) & 4.0($\pm$1.3) & 2.1($\pm$0.7)  \\
 
& & & & Lunar (1 pole) & 48 &	9.8 (+1.2/-1.4) & 4.3$(\pm$1.1) & 2.0($\pm$0.5) & 1.1($\pm$0.3)  \\
 &	  &    &   &    &   &	10.3 (+1.5/-1.8) & 4.7$(\pm$1.5) & 2.2($\pm$0.7) & 1.2($\pm$0.4)  \\

& & & & Lunar (2 pole) & 24 &	9.8 (+1.2/-1.4) & 4.3$(\pm$1.1) & 2.0($\pm$0.5) & 1.1($\pm$0.3)  \\
 &	  &    &   &    &   &	10.3 (+1.5/-1.8) & 4.7$(\pm$1.5) & 2.2($\pm$0.7) & 1.2($\pm$0.4)  \\

 \hline
50  &	0-0.06 &	0.03-0.20	& 0.36-0.63  &  L2 (Unrestricted FOR) & 60 &	15.9 (+1.1/-1.3) & 11.2$(\pm$1.7) & 5.2($\pm$0.8) & 2.8($\pm$0.4) \\
 &	  &    &   &    &   &	17.1 (+1.6/-1.8) & 13.0$(\pm$2.6) & 6.0($\pm$1.3) & 3.2($\pm$0.7)  \\

& & & & L2 (Restricted  FOR) & 60 &	20.3 (+1.6/-1.8) & 18.4$(\pm$3.2) & 8.5($\pm$1.4) & 4.6($\pm$0.7) \\
&	  &    &   &    &   &	22.0 (+2.1/-2.4) & 21.7$(\pm$4.7) & 10.0($\pm$2.1) & 5.3($\pm$1.1)  \\

& & & & Lunar (1 pole) & 120 & 15.9 (+1.1/-1.3) & 11.2$(\pm$1.7) & 5.2($\pm$0.8) & 2.8($\pm$0.4) \\
 &	  &    &   &    &   &	17.1 (+1.6/-1.8) & 13.0$(\pm$2.6) & 6.0($\pm$1.3) & 3.2($\pm$0.7)  \\

& & & & Lunar (2 pole) & 60 & 15.9 (+1.1/-1.3) & 11.2$(\pm$1.7) & 5.2($\pm$0.8) & 2.8($\pm$0.4) \\
 &	  &    &   &    &   &	17.1 (+1.6/-1.8) & 13.0$(\pm$2.6) & 6.0($\pm$1.3) & 3.2($\pm$0.7)  \\
 \hline
 
100  &	0-0.03 &	0.05-0.17	& 0.40-0.60 &  L2 (Unrestricted FOR) & 120 &	25.0 (+1.4/-1.6) & 28.5$(\pm$3.7) & 12.9($\pm$1.6) & 6.9($\pm$0.8)  \\
 &	  &    &   &    &   &	26.4 (+2.0/-2.1) & 32.2$(\pm$5.2) & 14.5($\pm$2.3) & 7.8($\pm$1.2)  \\

& & & & L2 (Restricted  FOR) & 120 &	29.6 (+1.9/-1.9) & 40.1$(\pm$5.3) & 18.2($\pm$2.3) & 9.8($\pm$1.2) \\
&	  &    &   &    &   &	31.8 (+2.7/-2.9) & 46.2$(\pm$7.6) & 21.0($\pm$3.5) & 11.2($\pm$1.9)  \\

& & & & Lunar (1 pole) & 240 &	25.0 (+1.4/-1.6) & 28.5$(\pm$3.7) & 12.9($\pm$1.6) & 6.9($\pm$0.8)  \\
 &	  &    &   &    &   &	26.4 (+2.0/-2.1) & 32.2$(\pm$5.2) & 14.5($\pm$2.3) & 7.8($\pm$1.2)  \\

& & & & Lunar (2 pole) & 120 &	25.0 (+1.4/-1.6) & 28.5$(\pm$3.7) & 12.9($\pm$1.6) & 6.9($\pm$0.8)  \\
 &	  &    &   &    &   &	26.4 (+2.0/-2.1) & 32.2$(\pm$5.2) & 14.5($\pm$2.3) & 7.8($\pm$1.2)  \\

\hline
\hline
\end{tabular}
\end{center}
\end{table*}

\section{Conclusions}

In this paper I discussed the occurrence rate of life-bearing planets, termed `the frequency of life': the goal of the DRAKE mission.  This is parameterised as \fL in the Drake and Seager equations.  It depends in turn on how the habitable zone itself is defined to give a subset of `candidate' planets.  In this study we have chosen the optimistic habitable zone limits of \cite{Kopparapu2013} but with other definitions, different results will be obtained.  There might also be planets where biosignature gases are not observable and yet life exists (e.g. in subsurface oceans). Since there is no information on the rate of such false negatives, these have been ignored in this study, as has the impact of the star on falsifying the detection of a true biosignature. Thus the observable \fL considered here may underestimate the true frequency of life, but to what extent is unknown.  Even with these limitations on the definition of \fL, an initial result from a spectroscopic survey would still deliver significant insights into how common life is and possibly the planetary basis for abiogenesis.

Using a bootstrap Monte Carlo simulation I explored how the uncertainty (95\% CI) on \fLo scales with number of planets in the sample. Given the assumption made here that a biosignature pattern can be correctly identified through observation at 3$\sigma$ significance, the simulation indicates that sampling error will dominate over this observational error.  The uncertainty is a function not only of the sample size, but also of \fLo itself: maximal at \fLo = 0.5, and minimal at \fLo = 0 or 1.  Even a completely negative sample (i.e. \fLo = 0) can constrain the upper limit of \fLt at 95\% confidence.  Due to diminishing return in terms of reducing the CI size with \Nsamp, a 50-planet sample size might be optimal as a target. If \fLo = 0, this would constrain \fLt to no more than 0.06 at 95\% confidence, and between 0.03-0.2 if \fLo = 0.1.  The statistical modeling approach presented in this paper can be further developed and refined in future studies.

The DRAKE mission is built around using transit spectroscopy to obtain atmospheric spectra and as such has a number of limitations compared to direct imaging approaches.  The first is that any sample will be inevitably biased in terms of stellar type.  This is because only M-dwarf HZ planets will be viable for this technique, so that extending the conclusions to Sun-like stars may not be possible.  Even among M-dwarfs, the sample may be biased towards early M-dwarfs with later types possibly under-represented due to their low brightness.  Compared to direct imaging approaches, the available sample sizes will be small and the efficiency of observing time to total mission time low.  However, transit spectroscopy is a mature technique that would require much less technological development compared to direct imaging in order to observe HZ planets.  Although the sample is confined to M-dwarfs, these could be considered a galactic norm representing 70\% of all stars, and therefore even if the mission is limited to finding \fL for this subset of systems, its conclusions will still be highly significant.   We have shown that small sample sizes can still provide  constraints on the 95\% CI for \fL and so using transit spectroscopy to find a first experimental estimate for \fL is a viable prospect. 

I presented a baseline observatory design and mission plan for DRAKE, and performed a feasibility study. The final results are very sensitive to the assumptions and detection criteria used in this study. $N_{\rm t}$ for each planet is highly sensitive to any factors that affect the calculation of $SNR_1$, as well as the choice of goal SNR for detection.  These include the choices made in instrument design, as well as star, planet and noise modelling. I find that such a mission is feasible under certain conditions and is therefore a viable alternative to direct imaging approaches. The concept warrants further study with more advanced instrumental design and noise modelling, together with performance metrics based on spectral retrievals.

In the current study, I conclude that for a 50-planet survey, 17- to 22m-class telescopes will most probably be required to achieve a mission of less than 10 years duration. The requirements depend on whether the putative observatory has an unrestricted or restricted FOR.  For a space-based design with a 50\% FOR, e.g. an observatory based at L2, I conclude that a 22-m telescope is likely to successfully achieve complete a 50-planet survey within 10 years. If a novel design permits almost unrestricted FOR at L2 or if we consider lunar polar telescopes, then the size of telescope can be reduced.  A 17-m telescope is likely to achieve the mission under such circumstances.

15-m class telescopes are unlikely to achieve a 50-planet survey in 10 years, however we show that on average a 23-planet survey can be achieved in the restricted FOR case, rising to 40 planets in the unrestricted FOR cases. Such surveys have the potential to constrain \fL to < 0.2 at 95\% confidence assuming 1 planet is positive in the survey.
Such a size of primary mirror has already been proposed for the Luvoir-A mission \citep{LUVOIR2019}. The Luvoir-A High Definition Imager (HDI) has the capability to perform transit spectroscopy in the wavelength range 200 nm to 2.5 \textmu m.  This wavelength range is shifted to shorter wavelengths than proposed for DRAKE, and so the feasibility study results obtained in this paper are  not directly transferable to Luvoir-A. However, the results obtained here pertaining to sample size and uncertainty on \fL are not exclusively applicable to transit spectroscopy of M-dwarfs or DRAKE, and could be more generally applied to other HZ surveys based either on transit spectroscopy or direct imaging (so long as definitive biosignature identification can be made).

Lunar-based polar telescopes could be considered for the DRAKE mission, and could utilise 17-m telescopes to achieve a 50-planet survey within 10 years. A single polar telescope would have an unchanging FOR, but to only half the sky.  Thus the unrestricted FOR simulations can be applied, but it must be assumed that at least twice as many planets must be known as in the survey.   The simulations also assumed that the initial pool of available planets for given survey, $N_{\rm {pool}}$ was 1.2$\times$ \Nsamp.  For the single lunar telescope this would mean at least 120 candidate planets (Earth-sized in the HZ of M-dwarfs) must be known. Given the likely yields for TESS and other missions in the coming decade, it seems unlikely at this time that 120 such planets  will become known in the next few years. From that standpoint the two-pole lunar telescope might be more feasible since that configuration has an $N_{\rm min}$ = 60.

These results are predictions based on Monte Carlo realisations of possible model planet populations. 
In reality much will depend on the actual final planet sample available at the time of the mission. Ultimately, the DRAKE mission, if given the go ahead, would adapt to the planet population known at the time.  This will shape the final expected mission time or the exact diameter of telescope needed.  We find here that simulations that assume an Earth-like mean molecular weight, $\mu$, for all planets (type 1) tend to underestimate the mission time slightly compared to simulations where $\mu$ was varied randomly (type 2), although in Table \ref{table: summary} most of the type 1 results are within 15\% of the type 2 results, and all are within 20\%.  It may be necessary to have some kind of adaptive scheduling during the real mission based on the likely cumulative SNR achieved for a given planet which will come to approximate the type 2 schedule simulated here.

Due to the low observing efficiency, I conclude that the DRAKE mission might be best incorporated into plans for a wider-ranging observatory. It could be a strong scientific goal for a future L2 or lunar observatory.  While the primary goal of DRAKE is to find \fL, it can also be used to survey other habitable zone planet characteristics on a statistical basis, e.g. water and CO$_2$ abundances as suggested by \cite{Bean2017}, providing experimental constraints to the habitable zone itself.
 
The discovery of the first life on another planet will have major scientific and cultural impacts.  However, elucidating how common life is, and the planetary conditions under which it arises, will allow us to probe even deeper questions into the occurrence and origin of life.  The DRAKE mission is a viable approach to attempt the first measurement of the frequency of life-bearing planets in the Cosmos.

\section*{Acknowledgements}
I acknowledge usage of the following software packages: NumPy v1.19.1 \citep{2020NumPy-Array}, SciPy v1.5.2 \citep{2020SciPy-NMeth}, Matplotlib v3.3.1 \citep{matplotlib}.  My thanks to Matt Griffin (Cardiff University) for his helpful comments and suggestions. I also thank the referee for their very helpful comments and feedback.

\section*{Data Availability Statement}
The data underlying this article will be shared on reasonable request to the corresponding author.




\bibliographystyle{mnras}
\bibliography{DRAKE.bib}

\bsp	
\label{lastpage}
\end{document}